\documentclass[prmaterials,psfig,pstricks,twocolumn,preprintnumbers,amsmath,amssymb,superscriptaddress]{revtex4}

\usepackage{graphicx}          
\usepackage{dcolumn}           
\usepackage{bm}                

\usepackage{array,booktabs,lmodern}
\usepackage[scaled=1.00]{helvet}
\usepackage[final]{listings,microtype}
\usepackage[T1]{fontenc}
\usepackage[version=3]{mhchem} 
\usepackage{siunitx} 
\usepackage[english]{babel}
\usepackage{float}
\usepackage{natbib}
\usepackage{setspace}
\usepackage{xkeyval}
\usepackage{natmove}
\usepackage{titlesec}

\newcommand{\mat}[1]{\mbox{\boldmath{$#1$}}} 
\usepackage{xfrac}
\usepackage{relsize}

\begin{document}

\preprint{Page}

\title{Investigating charge carrier scattering processes in anisotropic semiconductors through first-principles calculations: 
The case of p-type SnSe}

\author{Anderson S. Chaves}
\email{aschaves@ifi.unicamp.br}
\affiliation{Instituto de F\'{i}sica Gleb Wataghin, Universidade Estadual de Campinas, UNICAMP, 13083-859 Campinas, S\~{a}o Paulo, Brazil}
\author{Robert Luis Gonz\'{a}lez-Romero}
\email{robertl2703@gmail.com}
\affiliation{Departamento de Sistemas F\'{i}sicos, Qu\'{i}micos y Naturales, Universidad Pablo de Olavide, Ctra. de Utrera, km. 1, 41013 Sevilla, Spain}
\author{Juan J. Mel\'{e}ndez}
\email{melendez@unex.es}
\affiliation{Department of Physics, University of Extremadura, Avenida de Elvas, s/n, 06006 Badajoz,
and Institute for Advanced Scientific Computing of Extremadura (ICCAEx), Avda. de Elvas, s/n, 06006 Badajoz, Spain}
\author{Alex Antonelli}
\email{aantone@ifi.unicamp.br}
\affiliation{Instituto de F\'{i}sica Gleb Wataghin and Centre for Computational Engineering \& Sciences, Universidade Estadual de Campinas, UNICAMP, 13083-859 Campinas, S\~{a}o Paulo, Brazil}

\date{\today}

\begin{abstract}

Efficient \textit{ab initio} computational methods
for the calculation of thermoelectric transport properties of materials
are of great avail for energy harvesting technologies. 
The \texttt{BoltzTraP} code has been largely used to efficiently calculate thermoelectric coefficients. 
However, its current version that is publicly available is based only
on the constant relaxation time (RT) approximation, 
which usually does not hold for real materials.
Here, we extended the implementation of the \texttt{BoltzTraP} code by  
incorporating realistic k-dependent RT models 
of the temperature dependence of the main scattering processes, namely, screened polar and 
nonpolar scattering by optical phonons, scattering by acoustic phonons,
and scattering by ionized impurities with screening. 
Our RT models are based on a smooth Fourier interpolation of 
Kohn-Sham eigenvalues and its derivatives, taking into account 
non-parabolicity (beyond the parabolic or Kane models),
degeneracy and multiplicity of the energy bands on the same footing, 
within very low computational cost. 
In order to test our methodology, we calculated the
anisotropic thermoelectric transport properties 
of low temperature phase (\textit{Pnma}) of 
intrinsic p-type and hole-doped tin selenide (SnSe).
Our results are in quantitative agreement with experimental data, regarding 
the evolution of the anisotropic thermoelectric coefficients with both temperature and chemical potential. 
Hence, from this picture, we also obtained the evolution and understanding of the main scattering processes 
of the overall thermoelectric transport in p-type SnSe.

\end{abstract}


\maketitle

\section{Introduction}

Thermoelectric (TE) devices are heat engines based on 
an all-solid-state technology that attempt 
to harvest renewable energy from waste heat or 
function as heat pumps. Their properties make them suitable 
for several potential applications including 
automotive exhaust systems or solar energy
converters~\cite{bell2008cooling}. However, their widespread commercial 
use has been hampered by their low efficiency, and hence, 
efforts towards the enhancement of TE conversion efficiency has
been intensified in the last years.
Highly efficient TE devices demand materials that exhibit high figures of merit ($zT$), a dimensionless
quantity defined as $zT = (\sigma S^2/\kappa) T$ (where $S$ is the Seebeck coefficient, 
$\sigma$ is the electrical conductivity, $\kappa$ is the total thermal conductivity and $T$ is 
an average temperature of operation), along with a large temperature gradient across the materials.
Consequently, the optimization of $zT$ is obtained 
by lowering $\kappa$ while maintaining high the power factor ($PF = \sigma S^2$).

Several ways have been attempted in order to obtain optimized $zT$ values, 
including the reduction of $\kappa$ through the enhancement of 
scattering processes for phonons by embedding 
nanostructures~\cite{kanatzidis2009nanostructured,zhao2013high,johnsen2011nanostructures,vineis2010nanostructured}. 
In addition, other strategies deal with the challenge of 
maximizing $PF$, which is strongly hampered by the conundrum 
that $\sigma$ and $S$ are inversely interdependent and hence, 
recent strategies, based on modifications in the band structure 
of existing semiconductors have been proposed, such as increasing 
band degeneracy within convergence of bands~\cite{zhao2013high,liu2012convergence} or taking advantage 
of their band structure anisotropy~\cite{parker2015benefits} and non-parabolicity~\cite{chen2013importance}.

In order to evaluate TE transport coefficients,
the demand for more efficient computational approaches has increased. 
In particular, methods within the solution of Boltzmann transport equation (BTE)
and the relaxation time approximation (RTA) have been extensively used, 
including transport models largely based on the dispersive effective mass, $m_{eff}$, 
as an adjustable parameter 
in closed-form expressions for the scattering mechanisms. 
Those models, such as single parabolic band (SPB)~\cite{rode1971electron,bahk2014electron}       
or Kane models, 
adjust $m_{eff}$
to fit the calculated transport properties
to the experimental data and often impressively capture transport properties
over a range of temperatures and carrier concentrations.
Meanwhile, the anisotropy can be captured by the use of anisotropic effective mass tensors.
However, those models impose limitations in the understanding of TE properties and their 
improvement, since they are based on $m_{eff}$ and not on the density of states effective mass, $m^*$.
Even though within Kane models non-parabolicity can be taken into account, 
the consideration of multiplicity and degeneracy of band edges is not straightforward.
For example, in the two-band Kane model,
the interaction between valence band maximum (VBM) and conduction band minimum (CBM) 
is treated using the perturbative $\bf{k.p}$ method, while the
remaining bands are neglected.

Other BTE-RTA based models relying directly on the \textit{ab initio} bandstructure
have led to a more realistic description of TE transport properties.
Approximating the quasi-particle energies by Kohn-Sham (KS) eigenvalues,
within density functional theory (DFT) framework~\cite{hohenberg1964inhomogeneous,kohn1965self}, 
it is usually possible to access band structure 
dependent quantities with relative simplicity from the analytical representation of the bands
obtained from the interpolation 
schemes~\cite{chaput2005transport,hamann2009maximally,pickett1988smooth,uehara2000calculations,koelling1986interpolation}. 
The \texttt{BoltzTraP} code~\cite{madsen2006boltztrap} 
has been extensively adopted to this end~\cite{singh2008density,hautier2013identification,may2009influence,parker2013high,he2016ultralow},
particularly due to its numerical stability and efficiency in 
interpolating band structure by using the smooth Fourier interpolation method 
and then solving the linearized BTE-RTA by performing all the required integrations. 
However, the current version of \texttt{BoltzTraP} that is publicly available is based only
on the constant RTA, with resulting TE coefficients weighted by an unknown averaged relaxation time (RT), $\tau$.
This consideration ultimately takes $\tau$ to be direction independent, 
a feature that usually does not hold for real materials, which may possess
strongly anisotropic Fermi-surface topology. Additionally, $\tau$ 
may possesses a strong dependence with temperature and chemical potential that should
be considered for real materials.  

\textit{Ab initio} models for the scattering mechanisms and their RTs 
certainly improve the understanding
of transport properties and aid in the search 
for the next-generation of high-performing TE materials~\cite{ponce2020first}. 
In this context, electron-phonon (\textit{el-ph}) 
interactions have been fully treated within the \textit{ab initio}
framework, which enabled the calculation of mobility at different electron concentrations
for several systems, for example, silicon~\cite{restrepo2009first,qiu2015first,ponce2018towards}, 
SrTiO$_3$~\cite{himmetoglu2014first} and GaAs~\cite{zhou2016ab}.
However, to the best of our knowledge, such approaches have been limited to few
specific systems, which is mainly attributed to the high computational cost,
even in the case of short ranged \textit{el-ph} interactions of metals and nonpolar semiconductors.
For polar semiconductors and oxides, the situation is even more challenging from this viewpoint,
since the \textit{ab initio} treatment of the long range Fr{\"o}hlich interaction~\cite{rohlfing2000electron}
involves a very large number of \textit{e-ph} matrix elements to attain convergence. 
Recently, improvements regarding this problem have been
achieved by splitting up the \textit{e-ph} matrix elements into short- and long-range contributions~\cite{pellegrini2016physics,bostedt2016linac},
in which the long-range part can be treated by using an analytical formula based on the
Vogl model~\cite{martin2004electronic}, while the short-range part is well-behaved. Very recently,
a combination of the many-body theory of the electron-phonon interaction and the Boltzmann
transport formalism, within an \textit{ab initio} framework, was used to obtain the temperature-induced
renormalization of the electronic and transport problems of SnSe~\cite{Caruso2019,li2019resolving}.
However, the daunting task of calculating RTs over the entire Brillouin zone (BZ) is
still extremely computationally demanding for three-dimensional systems.

In this paper, our aim is to implement realistic 
models for the RT within \texttt{BoltzTraP},
that go beyond parabolic or Kane models, in order to include non-parabolicity, multiplicity and 
degeneracy of the band edges on the same footing, however, at a much lower computational cost 
than that of fully \textit{ab initio} approaches. 
In order to proceed, we considered ${\bf{k}}$-dependent RT models
of the temperature dependence of the main scattering processes, namely, screened polar and
nonpolar scattering by optical phonons, scattering by acoustic phonons,
and scattering by ionized impurities with screening.
Our implementation took advantage of the original \texttt{BoltzTraP} implementation 
concerning the calculations of band velocities within the RT formulae 
using the smooth Fourier interpolation of
KS eigenvalues and its derivatives. 
As it will be discussed later, our methodology can explain 
many experimental transport data using in fact only very few fitting parameters, 
derived for only one temperature and chemical potential and 
subsequently extending it to larger ranges of temperature and carrier concentration. 

In order to test our methodology, we calculated the
anisotropic TE transport properties
of the low temperature phase (\textit{Pnma}) of p-type tin selenide (SnSe), 
which is one of the highest-performing TE materials known that
has recently attracted much research 
interest~\cite{zhao2014ultralow,zhao2015ultrahigh,wu2017engineering,skelton2016anharmonicity,gonzalez2017insights,dewandre2016two}. 
The motivation for choosing this specific material comprises its intrinsic anisotropy along with 
non-parabolicity, degeneracy, and multiplicity of band edges, 
influencing TE transport properties.  
Hence, the understanding of their scattering mechanisms with temperature and chemical potential 
is of utmost importance in order to
obtain hints about next-generation high-performing TE materials. 
Consequently, the subject of the present research is to 
report on the dominant anisotropic charge carrier
scattering mechanisms in SnSe considering different
temperature and carrier concentration ranges, in which 
experimental transport data have been used
to probe the limits of our methodology. 

\section{Theoretical Approach}

\subsubsection{BTE within the RTA}

Our methodology for carrier transport is based 
on the numerical solution to the semiclassical BTE, which
has been applied to some extent to semiconductors. 
In the diffusive transport limit, in the presence of a 
temperature gradient ($\nabla T$) and applied electric (${\bf{E}}$) 
and magnetic fields (${\bf{B}}$), 
carrier transport properties can be obtained
by solving the semiclassical BTE for the nonequilibrium 
carrier distribution function $f_{n,{\bf{k}}} = f(\epsilon_{n,{\bf{k}}})$
\begin{equation}
\label{boltz1}
\frac{\partial f_{n,{\bf{k}}}}{\partial t} + {\bf{v}}_{n,{\bf{k}}}\cdot\nabla_{{\bf{r}}} f_{n,{\bf{k}}} - \frac{{\bf{F}}}{\hbar}\cdot\nabla_{{\bf{k}}} f_{n,{\bf{k}}} = \left(\frac{\partial f_{n,{\bf{k}}}}{\partial t}\right)_{coll}~, 
\end{equation}
where, $n$ and ${\bf{k}}$ label the band index and the wavevector, respectively. The
electronic band velocity of the carrier in the $\{n,{\bf{k}}\}$ state with energy $\epsilon_{n,{\bf{k}}}$ 
is defined as ${\bf{v}}_{n,{\bf{k}}} = 1 / {\hbar} \nabla_{{\bf{k}}}\epsilon_{n,{\bf{k}}}$,
while the external force is given by ${\bf{F}} = e({\bf{E}} + {\bf{v}}_{n,{\bf{k}}}\times{\bf{B}})$, 
where $e$ is the absolute value of the charge carrier.
The right-hand side term is related to charge carrier collisions and
describes different sources of scattering and dissipation
driving the system into a steady state. This scattering term can be expressed
by introducing the per-unit-time probability, $W(n,{\bf{k}}|j,{\bf{{k}^{\prime}}})$, of the 
transition of the charge carrier from the state $\{n,{\bf{k}}\}$ to
state $\{j,{\bf{{k}^{\prime}}}\}$, as a result of a particular scattering mechanism.
From the principle of detailed equilibrium, the number of charge carriers
coming into the state $\{j,{\bf{{k}^{\prime}}}\}$ from $\{n,{\bf{k}}\}$ is the same as
the number coming out from $\{j,{\bf{{k}^{\prime}}}\}$ into $\{n,{\bf{k}}\}$, we have
\begin{multline}
\label{boltz2}
\left(\frac{\partial f_{n,{\bf{k}}}}{\partial t}\right)_{coll} = \sum_{j,{\bf{{k}^{\prime}}}} [W(j,{\bf{{k}^{\prime}}} | n,{\bf{k}}) f_{j,{\bf{{k}^{\prime}}}} \left(1 - f_{n,{\bf{k}}}\right) \\
- W(n,{\bf{k}} | j,{\bf{{k}^{\prime}}}) f_{n,{\bf{k}}} \left(1 - f_{j,{\bf{{k}^{\prime}}}}\right)]~.
\end{multline}
The knowledge of $f_{n,{\bf{k}}}$ allows the evaluation of the charge current density 
\begin{multline}
{\bf{j}} = -\frac{2e}{V}\sum_n\sum_{k}{\bf{v}}_{n,{\bf{k}}}f_{n,{\bf{k}}} = -\frac{2e}{(2\pi)^3}\sum_n\int{{\bf{v}}_{n,{\bf{k}}}f_{n,{\bf{k}}}d{\bf{k}}}~,
\end{multline}
and the heat energy flux density
\begin{multline}
{\bf{j}}_Q = \frac{2}{V}\sum_n\sum_{{\bf{k}}}\left(\epsilon_{n,{\bf{k}}}-\mu\right){\bf{v}}_{n,{\bf{k}}}f_{n,{\bf{k}}} \\
= \frac{2}{(2\pi)^3}\sum_n\int{\left(\epsilon_{n,{\bf{k}}}-\mu\right){\bf{v}}_{n,{\bf{k}}}f_{n,{\bf{k}}}d{\bf{k}}}~,
\end{multline}
where the factor $\num{2}$ appears due to the electron
spin, $V$ is the crystal's volume, and $\mu$ is the chemical potential. 

In order to proceed, we consider that the system is close enough to equilibrium so that
the nonequilibrium distribution function, $f_{n,{\bf{k}}}$, differs only slightly 
from that of the equilibrium state, $f_{n,{\bf{k}}}^{(0)}$, that is 
$\Delta f(n,{\bf{k}}) = \lvert{f_{n,{\bf{k}}}-f_{n,{\bf{k}}}^{(0)}}\lvert << f_{n,{\bf{k}}}^{(0)}$. 
Within such approximation, Eq.~\eqref{boltz2} can be written in the RTA
\begin{equation}
\label{coll}
\left(\frac{\partial f_{n,{\bf{k}}}}{\partial t}\right)_{coll} = -\frac{\Delta f(n,{\bf{k}})}{\tau_{n,{\bf{k}}}}~, 
\end{equation}
where
\begin{multline}
\label{tau1}
\frac{1}{{\tau_{n,{\bf{k}}}}} = \sum_{{\bf{{k}^{\prime}}}} \sum_j W(n,{\bf{k}}|j,{\bf{{k}^{\prime}}}) \\
\left( \frac{1-f^{(0)}_{j,{\bf{{k}^{\prime}}}}}{1-f^{(0)}_{n,{\bf{k}}}} - \frac{f^{(0)}_{n,{\bf{k}}}}{f^{(0)}_{j,{\bf{{k}^{\prime}}}}}\frac{\Delta f(j,{\bf{{k}^{\prime}}})}{\Delta f(n,{\bf{k}})}\right)~, 
\end{multline}
given the absence of quantization effects, thus $W(n,{\bf{k}}|j,{\bf{{k}^{\prime}}})$ does not depend on 
${\bf{E}}$, ${\bf{B}}$, or $\nabla T$. 
Consequently, in the case of a homogeneous system, zero magnetic field, and a time-independent
electric field, in the steady-state limit, Eq.~\eqref{boltz1} simplifies to
\begin{equation}
\label{boltz3}
{\bf{v}}_{n,{\bf{k}}}\cdot\nabla_{{\bf{r}}} f_{n,{\bf{k}}} - \frac{e {\bf{E}}}{\hbar}\cdot\nabla_{{\bf{k}}} f_{n,{\bf{k}}} = -\frac{\Delta f(n,{\bf{k}})}{\tau_{n,{\bf{k}}}}~, 
\end{equation}
from which it is possible to obtain the nonequilibrium 
distribution function provided that $\tau_{n,{\bf{k}}}$ does not depend on 
${\bf{E}}$ or $\nabla T$. 

Moreover, from the smallness of the
deviation from equilibrium, $f_{n,{\bf{k}}}$ can be expanded to first order as 
\begin{equation}
\label{smallness}
f_{n,{\bf{k}}} = f_{n,{\bf{k}}}^{(0)} -{\tau}_{n,{\bf{k}}}{\bf{v}}_{n,{\bf{k}}}\cdot{\bf{\Phi_0}}(\epsilon)\left(\frac{\partial f^{(0)}}{\partial \epsilon}\right)~,
\end{equation}
where ${\bf{\Phi_0}}(\epsilon) = -e{\mathlarger{\mat{\varepsilon}}} - \frac{\epsilon - \mu}{T}\nabla T$ is 
the generalized disturbing force (dynamic and static) causing the deviation from the
equilibrium distribution, where ${\mathlarger{\mat{\varepsilon}}} = {\bf{E}} + (1/e)\nabla \mu = -\nabla (\phi_0-(\mu/e))$ 
is the gradient of the electrochemical potential. 
Substituting Eq.~\ref{smallness} into Eq.~\eqref{tau1} we obtain
\begin{multline}
\label{tau2}
\frac{1}{\tau_{n,{\bf{k}}}} = \sum_{{\bf{{k}^{\prime}}}} \sum_j W(n,{\bf{k}}|j,{\bf{{k}^{\prime}}}) \\
\frac{1-f^{(0)}(\epsilon^{\prime})}{1-f^{(0)}(\epsilon)}\left( 1 - \frac{{\tau}_{j,{\bf{{k}^{\prime}}}}}{\tau_{n,{\bf{k}}}} \frac{{\bf{v}}_{j,{\bf{{k}^{\prime}}}}\cdot\bf{\Phi_0}(\epsilon^{\prime})}{{\bf{v}}_{n,{\bf{k}}}\cdot\bf{\Phi_0}(\epsilon)}\right)~.
\end{multline}
Considering that the per-unit-time probability of charge carrier transition, $W(n,{\bf{k}}|j,{\bf{{k}^{\prime}}})$, 
does not depend on ${\bf{k}}$ and ${\bf{{k}^{\prime}}}$ separately, but only on the magnitudes of the vectors, $\lvert{\bf{k}}\lvert$ and
$\lvert{\bf{{k}^{\prime}}}\lvert$, and the angle between them, {${\bf{k}}\cdot{\bf{{k}^{\prime}}}$}, that is
$W(n,{\bf{k}}|j,{\bf{{k}^{\prime}}}) = W_{n,j}(\lvert{\bf{k}}\lvert,\lvert{\bf{{k}^{\prime}}}\lvert,{\bf{k}}\cdot{\bf{{k}^{\prime}}})$.
Additionally, considering that the dispersion relation is an arbitrary 
spherically symmetric function of the magnitude of the 
wavevector, $k = \lvert{\bf{k}}\lvert$,(not necessarily parabolic) and the charge carrier 
scattering is purely elastic, that is, 
the charge carriers exchange energy during scattering only by 
impulses with $\epsilon(\lvert{\bf{k}}\lvert)=\epsilon(\lvert{\bf{{k}^{\prime}}}\lvert)$, Eq.~\eqref{tau2} 
may be rewritten as~\cite{askerov2009thermodynamics}
\begin{equation}
\label{tau3}
\frac{1}{{\tau_{n,{k}}}} = \sum_{{\bf{{k}^{\prime}}}} \sum_j W(n,{\bf{k}}|j,{\bf{{k}^{\prime}}}) \left(1-\frac{{\bf{k}}\cdot{\bf{{k}^{\prime}}}}{k^2}\right)~.
\end{equation}
At this point, some comments are in order. Although Eq.~\eqref{tau3} has been derived for isotropic bands, results obtained from it 
have been used to study transport properties of lead chalcogenides, which are anisotropic~\cite{ahmad2010energy,ravich1971scattering}.
The use of this methodology in this case is possible because, although anisotropic, the transport properties in these materials along the different directions are mutually independent, except for the magnetoresistance, which essentially depends on the anisotropy. As it will be clear later, the application of this methodology in the case of SnSe is justified by the good agreement of our results with the experiment at different values of temperature and chemical potential.

\subsubsection{TE Kinetic Coefficients Tensors}

The tensorial formalism is appropriate to discuss TE effects 
for anisotropic materials. 
The off-diagonal coupling between the electronic current density, ${\bf{j}}$, and 
heat energy flux density, ${\bf{j}}_Q$, can be derived 
on a more general basis within the linear response theory~\cite{callen1985thermodynamics} as
\begin{equation}
\begin{bmatrix} 
    {\bf{j}} \\ 
    {\bf{j}}_Q  
\end{bmatrix}
=
\begin{bmatrix}
    {\bf{L^{11}}} & {\bf{L^{12}}} \\
    {\bf{L^{21}}} & {\bf{L^{22}}}
\end{bmatrix}
\cdot
\begin{bmatrix} 
    {\mathlarger{\mat{\varepsilon}}} \\
    -\frac{\nabla T}{T}
\end{bmatrix}
\end{equation}
in which, ${\bf{L^{11}}}$, ${\bf{L^{12}}}$, ${\bf{L^{21}}}$,
${\bf{L^{22}}}$ are the moments of the generalized transport 
coefficients, which we will call kinetic coefficient tensors, with 
${\bf{L^{12}}} = {\bf{L^{21}}}$ based on the Onsager 
reciprocity relations~\cite{callen1985thermodynamics}.
Such kinetic coefficients can be expressed as 
\begin{equation}
\label{Lambda}
\Lambda^{(\alpha)}(\mu;T)= e{^2}\int\Xi(\epsilon,\mu,T)(\epsilon - \mu)^{\alpha}\left(-\frac{\partial f^{(0)}(\mu;\epsilon,T)}{\partial \epsilon}\right)d\epsilon~, 
\end{equation}
with ${\bf{L^{11}}} = \Lambda^{(0)}$, 
${\bf{L^{21}}} = {\bf{L^{12}}} = -(1/e)\Lambda^{(1)}$, and 
${\bf{L^{22}}} = (1/e^2)\Lambda^{(2)}$, in which $\Xi(\epsilon,\mu,T)$ 
is the transport distribution kernel (TDK) given by
\begin{equation}
\Xi(\epsilon,\mu,T) = \int \sum_n{{\bf{v}}_{n,{\bf{k}}}\otimes{\bf{v}}_{n,{\bf{k}}}{\tau}_{n,{k}}}(\mu,T)\delta(\epsilon - \epsilon_{n,{\bf{k}}})\frac{d{\bf{k}}}{8\pi^3}~. 
\end{equation}
At both experimental conditions of zero temperature gradient ($\nabla T = 0$) 
and zero electric current, the kinetic coefficient tensors 
can be identified with the electrical conductivity tensor, $\sigma = \Lambda^{(0)}$, 
with the Seebeck coefficient tensor, $S = (eT)^{-1}\Lambda^{(1)}/\Lambda^{(0)}$, 
and with the charge carrier contribution to the thermal conductivity tensor, 
$\kappa_{elec} = (e^2T)^{-1} \left({\Lambda^{(1)}\cdot{\Lambda^{(0)}}^{-1}}\cdot{\Lambda^{(1)}} - \Lambda^{(2)}\right)$. 

\subsection{Scattering Mechanisms and the RTA}\label{mechanisms}

Within the Born approximation in the scattering theory, the magnitude of the Hamiltonian 
of the charge carrier interaction, $H^{\prime}$, is considered to deviate 
only slightly from the magnitude of the non-perturbed Hamiltonian, $H$, that is, $H^{\prime} - H << H$. 
The transition probability per-unit-time between Bloch states $\Psi_{j,{\bf{{k}^{\prime}}}}$ and $\Psi_{n,{\bf{k}}}$ 
can be obtained from the first order perturbation theory (Fermi's golden rule)
\begin{equation}
\label{1PT}
W(n,{\bf{k}}|j,{\bf{{k}^{\prime}}}) = \frac{2\pi}{\hbar}\lvert\langle\Psi_{j,{\bf{{k}^{\prime}}}}\lvert{H^{\prime}}\lvert\Psi_{n,{\bf{k}}}\rangle\lvert^2\delta(\epsilon_{j,{\bf{{k}^{\prime}}}}-\epsilon_{n,{\bf{k}}})~,
\end{equation}
which is valid in the so-called weak coupling regime~\cite{hameau1999strong}.
Integrating out such probability over the BZ, the total scattering rate is obtained.
Specifically, in the presence of a phonon field, an electron in the Bloch state $\Psi_{n,{\bf{k}}}$ will 
experience a perturbation $H^{\prime}$, inducing a transition to the state $\Psi_{j,{\bf{k}^{\prime}}}$. 
In such a process, momentum and energy are conserved so that $\bf{{k}^{\prime}} = \bf{k} \pm \bf{q}$ 
and $\epsilon_{j,{\bf{{k}^{\prime}}}}=\epsilon_{n,{\bf{k}}} \pm \hbar \omega_{{\bf{q}}}^{\lambda}$, 
where $\omega_{{\bf{q}}}^{\lambda}$ is the phonon frequency 
with wave vector $\bf{q}$ and mode number $\lambda$. 
The plus-minus sign refers to phonon absorption or emission, respectively. 
The respective per-unit-time transition probability is calculated from Eq.~\eqref{1PT} as 
\begin{multline}
\label{2PT}
W(n,{\bf{k}}|j,{\bf{k}} \pm {\bf{q}}) = \frac{2\pi}{\hbar}\lvert\langle\Psi_{j,{\bf{k}} \pm {\bf{q}}}\lvert{{H^{\prime}}_{{\bf{q}}}^{\lambda}}\lvert\Psi_{n,{\bf{k}}}\rangle\lvert^2 \\
\delta(\epsilon_{j,{\bf{k}} \pm {\bf{q}}}-\epsilon_{n,{\bf{k}}} \mp \hbar\omega_{{\bf{q}}}^{\lambda})~.
\end{multline}
By using the above equations along with equation~(\ref{tau3}), 
expressions for the different scattering 
mechanisms RTs can be derived, mostly following Ref.~\cite{askerov2009thermodynamics}, 
as will be discussed below.   

\subsubsection{Carrier-Acoustic Phonons Non-polar Scattering}

The deformation potential technique, as introduced by Bardeen and
Shockley~\cite{bardeen1950deformation} and extended by Herring and Vogt~\cite{herring1956transport} 
has been used to derive an expression
for the RT for the non-polar scattering of charge carrier by 
acoustic phonons. 
When an acoustic wave with vanishing $\bf{q}$ vector travels through a 
finite crystal it may induce shifts in the spacing
between neighboring atoms, resulting in local fluctuations of the energy band gap.
These are known as acoustic deformation potentials (ADPs), which can be measured
by quantifying the energy variation of the valence and conduction band edges per unit of strain.
The former variation represents the interaction energy of holes with lattice oscillations. 
The magnitude of the shifts, $\bf{{u}_s}$, are given in terms of plane waves as
\begin{multline} 
\label{shift1}
{\bf{u_{s}(r)}} = \frac{1}{\sqrt{N}} \sum_{{\bf{q}}} \sum_{\lambda = 1}^3 {\bf{e}}_{\lambda}({\bf{q}})\cdot \\
\left[b_{\lambda,{\bf{q}}}\exp{(i{\bf{q}}\cdot{\bf{r}})} + b_{\lambda,{\bf{q}}}^*\exp{(-i{\bf{q}}\cdot{\bf{r}})}\right]~, 
\end{multline}
where $\bf{r}$ are the atomic coordinates in real space, 
$N$ is the number of atoms in the periodic crystal,  
${\bf{e}}_{\lambda}({\bf{q}})$ is the polarization unit vector and  
$b_{\lambda,\bf{q}}$ are the complex normal coordinates. 
In Eq.~\eqref{shift1}, all atoms in the elementary 
unit cell oscillate in phase 
and the interaction energy must be proportional to the first 
derivative of the ${\it{\bf{{u}_{s}(r)}}}$ with respect 
to $\bf{r}$
\begin{equation}
H^{\prime}_{ac} = E_1 \nabla{\bf{{u}_{s} (r)}}~,
\end{equation}
where $E_1$ is the effective deformation potential. 

Going back to Eq.~\eqref{1PT}, it is possible to obtain the per-unit-time 
transition probability, and, along with Eq.~\eqref{tau3}, the RT 
for a given electronic band $n$ 
can be written as 
\begin{equation}
\label{tau_ac_0}
\frac{1}{\tau_{\bf{k}}} = \frac{2\pi}{M N}\frac{E_1^2}{\hbar v_0^2}k_B T \sum_{{\bf{k}}^{\prime}} \left(1-\frac{{\bf{k}}\cdot{\bf{{k}^{\prime}}}}{k k^{\prime}}\right) \delta\left(\epsilon_{{\bf{k}}^{\prime}} - \epsilon_{{\bf{k}}}\right)~,
\end{equation}
where $M$ is the atomic mass and $v_0$ is the sound velocity. 
In Eq.~\eqref{tau_ac_0}, it was considered that the scattering is elastic, and hence, 
the phonon energy has been neglected. Such approximation, 
is valid for $T >> \num{1}$K~\cite{askerov2009thermodynamics}, which meets 
the conditions usually found in TE applications. 
Finally, considering that the electronic dispersion is 
arbitrary but spherically symmetric, the summation over $\bf{{k}^{\prime}}$ 
can be turned into an integral that can be solved by using spherical coordinates and 
properties of $\delta$-functions. Hence, one finds 
the following expression for the RT for 
non-polar scattering of charge carriers by acoustic phonons in a single band $n$
\begin{equation}
\label{tau_ac}
\tau_{ac}(k) = \frac{\pi\hbar\rho}{E{_{1}}^2}\frac{v{_0}^2}{k{_B}T}\frac{1}{k^2}\left\vert\frac{\partial \epsilon_k}{\partial k}\right\vert~,
\end{equation}
where, $\rho = MN/V$ is the mass 
density of the material and $V$ is the crystal volume.

\subsubsection{Carrier-Optical Phonons Non-polar Scattering}

In more complex crystal lattices with 
two or more atoms in the unit cell, alongside 
with the scattering by acoustic phonons, the 
polar and non-polar scattering by optical phonons are at play. 
The idea of deformation potentials has been extended to the interaction between charge carriers and
long-wavelength (${\bf{q}}\rightarrow 0$) non-polar optical phonons, 
giving rise to the optical deformation potentials (ODP), corresponding
to shifts of the electronic bands due to relative displacement
between two sublattices of the crystal.
Unlike the scattering by acoustic lattice oscillations, optical phonons 
oscillate out of phase, and the center of mass remains at rest. 
In the case of the scattering by non-polar 
optical phonons, the interaction energy related to the induced variation of the energy band gap 
should be proportional to the shift of any atom in the unit cell~\cite{askerov2009thermodynamics}
\begin{equation}
\label{np_opt}
H^{\prime}_{npol} = \sum_{\lambda=4}^{3s}{\bf{A}}_{\lambda}{\bf{{u}_s}}^{\lambda}~,
\end{equation}
where ${\bf{A}}_{\lambda}$ is a constant vector related to the 
symmetry of the arrangement of the band edges, ${\bf{{u}_s}}^{\lambda}$ 
is the atomic displacement (similar to Eq.~\eqref{shift1}) associated with mode $\lambda$, and 
$s$ is the number of atoms in the unit cell. Consequently, the energy operator 
in this case appears to be much more involving than the case of acoustic phonons. 
In order to proceed, an effective vector can be taken to represent all interactions in an average manner, 
and hence, all the complexity can be encompassed within such effective vector.
Considering a specific optical phonon branch, $\lambda$, in 
which the scattering is occurring on its minimum, $\bf{A}$ can be defined as 
\begin{equation}
{\bf{A}} = E_0 {\bf{b}}_g~,
\end{equation} 
where $E_0$ is the effective ODP,  
${\bf{b}}_g = (\pi/a){\bf{g}}$ is the reciprocal lattice vector 
with $\bf{g}$ being a unit vector 
directed from the BZ center to the minimum, 
while $a$ is the lattice constant. 

Within the previous considerations, the per-unit-time transition 
probability for charge carriers, for a specific band $n$, can be obtained as
\begin{multline}
\label{W_npol}
W_{npol}({\bf{k}}|{\bf{{k}^{\prime}}}) = \frac{\pi E_0^2}{N M \omega_0} \left(\frac{\pi}{a}\right)^2 [N_0 \delta(\epsilon_{{k}^{\prime}} - \epsilon_k - \hbar \omega_{{\bf{q}}}) \\
+ (N_0+1)\delta(\epsilon_{{k}^{\prime}} - \epsilon_k + \hbar \omega_{{\bf{q}}})]~,
\end{multline}
with $\omega_0 = \omega_{({\bf{q}}\rightarrow 0)}$ and 
$N_0$ is the Bose-Einstein distribution.
Additionally, we considered only coupling parallel to 
the unit polarization vector (${\bf{e}}_{\lambda}({\bf{q}})$).
The non-polar scattering of charge carriers by optical phonons is essentially non-elastic, 
due to the magnitude of optical phonon energy. However, the description of this scattering 
within the RT approximation may arise from the evenness of the transition probability 
function, $W({\bf{k}}|{\bf{{k}^{\prime}}}) = W(|{\bf{k}}-{\bf{{k}^{\prime}}|})$, which is 
a condition satisfied by the scattering probability as given by Eq.~\eqref{W_npol}. 
On the basis of such considerations, 
at high temperatures, $k_BT >> \hbar\omega_0$, the following simple expression is obtained for
the RT for the non-polar scattering by optical phonons at a specified band $n$~\cite{askerov2009thermodynamics}
\begin{equation}
\label{tau_npol}
\tau_{npol}(k) = \frac{1}{\pi\hbar}\beta^2 \frac{{\rho}a^2}{k_{B}T}\frac{1}{k^2}\left\vert\frac{\partial \epsilon_k}{\partial k}\right\vert~,  
\end{equation}
in which $\beta={\hbar\omega_0}/{E_0}$ is the reciprocal ODP normalized by the energy of optical phonons.
In particular, $\tau_{npol}(k)$ presents the same temperature dependence as 
the scattering by acoustic phonons.

\subsubsection{Carrier-Optical Phonons Polar Scattering}

In semiconductor compounds with some degree of ionic bonding, 
there is an additional interaction with charge carriers known as
polar mode scattering, in which the charge carriers are scattered by the electric polarization
caused by longitudinal optical (LO) phonons. This was firstly discussed
by Fr{\"o}hlich~\cite{frohlich1937h} and Callen~\cite{callen1949electric},
while Howarth and Sondheimer~\cite{howarth1953theory} developed the theory of polar mode scattering,
on the basis of electrons as charge carriers on a simple parabolic
conduction band.
Following the derivation by Fr{\"o}hlich~\cite{frohlich1954electrons}, 
the electric polarization due to 
ion displacement is given by
\begin{multline}
\label{pol}
{\bf{P}}({\bf{r}}) = \left(\frac{\hbar\omega^{LO}(q)}{8\pi V \zeta^{\star}}\right)^{1/2} \sum_{{\bf{q}}} \sum_{\lambda=4}^{3s} {\bf{e}}_{\lambda}\cdot \\
\left[b_{\lambda}({\bf{q}}) \exp{(i {\bf{q}}\cdot{\bf{r}})} + b_{\lambda}^{*}({\bf{q}}) \exp{(-i{\bf{q}}\cdot{\bf{r}})}\right]~,
\end{multline}
where $\omega^{LO}(q)$ is the frequency of the LO phonons; 
$1/{\zeta^{\star}} = 1/{\zeta_{\infty}} - 1/{\zeta_0}$, with $\zeta_{\infty}$ and $\zeta_0$ being the 
high-frequency and the static dielectric constants, respectively.
From the Poisson's equation we can derive the scalar potential of the polarization vector, 
$\nabla^{2} \phi = 4\pi \nabla \cdot {\bf{P}}({\bf{r}})$, which yields the following 
interaction energy of the polar mode scattering of charge carriers by 
optical phonons
\begin{multline}
\label{H_pol}
H^{\prime}_{pol} = \pm e\phi = \mp i e \left(\frac{4\pi \hbar \omega^{LO}(q)}{2V\zeta^{*}}\right)^{1/2} \sum_{{\bf{q}}} \sum_{\lambda=4}^{3s} \frac{1}{q^2}({\bf{e}}_{\lambda}\cdot{\bf{q}}) \\
\left[b_{\lambda}({\bf{q}}) \exp{(i{\bf{q}}\cdot{\bf{r}})} + b_{\lambda}^{*}({\bf{q}}) \exp{(-i{\bf{q}}\cdot{\bf{r}})}\right]~,
\end{multline}
with ${\bf{e}}_{\lambda}\cdot{\bf{q}}=q$ in the case of LO phonons.

If the phonon dispersion is not considered, $\omega^{LO}(q)=\omega_0^{LO}$, then 
the following expression for the transition probability, at a specific band $n$, is obtained
\begin{multline}
\label{Wpol}
W_{pol}({\bf{k}}|{\bf{k}}^{\prime}) = \frac{4\pi^2e^2}{V \zeta^{\star}}\frac{\omega_0^{LO}}{({\bf{{k}^{\prime}}}-{\bf{k}})^2} \times\\
\left[N_0\delta(\epsilon_{{\bf{{k}^{\prime}}}}-\epsilon_{{\bf{k}}}+\hbar\omega_0^{LO}) + (N_0 + 1)\delta(\epsilon_{{\bf{{k}^{\prime}}}}-\epsilon_{{\bf{k}}}+\hbar\omega_0^{LO})\right]~.
\end{multline} 
Unlike non-polar mode scattering, from Eq.~\eqref{Wpol} it can be 
seen that $W_{pol}({\bf{k}}|{\bf{{k}^{\prime}}})$
is dependent on the directions ${\bf{k}}$ and ${\bf{{k}^{\prime}}}$ and, generally, it is impossible 
to introduce RT for polar mode scattering of optical phonons. 
However, at high temperatures, $k_BT >> \hbar\omega_0^{LO}$, the non-elasticity can be neglected with  
$\epsilon_{{\bf{{k}^{\prime}}}} \sim \epsilon_{{\bf{k}}}$ and $N_0 + 1 \sim N_0 \sim k_BT/\hbar\omega_0^{LO}$, 
which yields
\begin{equation}
\label{Wpol2}
W_{pol}({\bf{k}}|{\bf{{k}^{\prime}}}) = \frac{8\pi^2e^2}{V \zeta^{\star}}\frac{k_BT}{\hbar} \frac{1}{({\bf{{k}^{\prime}}}-{\bf{k}})^2}\delta(\epsilon_{{\bf{{k}^{\prime}}}}-\epsilon_{{\bf{k}}})~.
\end{equation}
Thus, substituting Eq.~\ref{Wpol2} into Eq.~\eqref{tau3} and transforming the summation over $\bf{{k}^{\prime}}$ 
into an integral that can be solved by using spherical coordinates and 
properties of $\delta$-functions, it results in the following closed expression 
for the RT of polar mode scattering of optical phonons, given a specific band $n$, 
\begin{equation}
\label{tau_pol}
\tau_{pol}(k) = \frac{\zeta^{\star}\hbar}{2{e^2}k{_B}T}\left\vert\frac{\partial \epsilon_{k}}{\partial k}\right\vert~,  
\end{equation}
which does not depend on the phonon-frequency in the high temperature limit. 

The generalization to include 
screening effects was subsequently developed by Ehrenreich~\cite{ehrenreich1959screening}. 
Free carriers that are present in the sample screen out 
the electric field produced by optical vibrations, resulting
twofold effects in the quasi-static approximation,
namely, a change in the matrix element of charge carrier interaction
with optical phonons and a shift in the frequency of
longitudinal optical mode~\cite{ehrenreich1959screening}.
The former effect decreases the transition probability by a factor of $1-(r_0q)^{-2}$,
in which $r_0$ is the screening radius given by
\begin{equation}
\label{r0}
r{_0}^{-2}(k) = \frac{4\pi{e^2}}{\zeta{_0}}\int{-\frac{\partial f^{(0)}(\epsilon)}{\partial \epsilon_k}}g(\epsilon)d\epsilon~, 
\end{equation}
where $f^{(0)}(\epsilon)$ is the equilibrium electron
distribution function and $g(\epsilon)$ is the density of states (DOS), which is given explicitly in Eq.~\eqref{dos2}.
The latter effect leads to a frequency shift of the
LO vibrations given by
\begin{equation}
\label{freq_shift}
{(\omega^{LO}})^2 = {(\omega^{TO}})^2\left(\frac{\zeta_0/\zeta_{\infty}+ (r_0q)^{-2}}{1+(r_0q)^{-2}}\right)~,
\end{equation}
where $\omega^{TO}$ is the transverse optical (TO) mode frequency. 
Therefore, the frequency of the LO phonon
is strongly reduced, which also affects the transition probability~\cite{ravich1971scattering}.
The dynamical features of the screening were neglected here, since their effect is
regarded  to be quite small~\cite{ravich1971scattering}.
Hence, the consideration of quasi-static screening results in the following factor 
in the denominator of expression for the RT
\begin{equation}
\label{screen_pol}
F_{pol} = \left[1 -\frac{1}{2(r_0k)^2}\ln[1+4(r_0k)^2]+\frac{1}{1+4(r_0k)^2}\right]^{-1}~. 
\end{equation}
It is important to note that the energy dependence of RT
is also changed because of the energy dependence of the screening through $r_0$~\cite{ravich1971scattering}.

The necessary modifications when dealing with 
polar mode scattering of p-like symmetry holes have also been 
addressed~\cite{kranzer1972polar,costato1972hole,kranzer1973hall,wiley1971polar}, 
beyond the polar mode scattering of s-like electrons.
The work by Wiley~\cite{wiley1971polar}
gave the first quantitative discussion of overlap effects 
on the polar mobility of holes, showing that, for carriers with p-like valence bands, 
the mobility is about twice the mobility for carriers with pure s-like wave functions.
Conclusions, in the same line of reasoning, that the mobility 
increases due to the p-like symmetry of wave functions, 
were obtained by Kranzer~\cite{kranzer1972polar} on the basis of 
a numerical solution of the BTE
and by Costato \textit{et al.}~\cite{costato1972hole}
by using a Monte Carlo technique for solving the coupled BTE.
Consequently, besides screening effects, it is also necessary 
to include the correction factor, $K_{pol}$, 
in the RT given by Eq. \eqref{Tau_pol}, 
due to the p-like symmetry of the wave functions,  
which is important, for example, in the case of polar mode scattering of holes. The way the correction
factor $K_{pol}$ is determined in the actual calculations will be discussed later in the article.

\subsubsection{Carrier-Impurity Scattering}

The consideration of extrinsic collision processes,
beyond those involving the crystalline lattice, i.e., the intrinsic
scattering by phonons, requires the presence of impurities in the crystal.
Particularly, we will consider only the concentration of ionized impurities
because the number of charged donors or acceptors is usually considerably larger
than that of neutral imperfections. 
Ionized impurity scattering has been treated theoretically by 
Brooks and Herring (B-H)~\cite{brooks1955theory,chattopadhyay1981electron}, by considering a screened Coulomb potential, the
Born approximation for the evaluation of transition probabilities,
and neglecting the perturbation effects of the impurities on the electron energy levels and wave functions.
In the B-H theory, the electron is scattered independently
by dilute concentrations of ionized centers randomly distributed in semiconductors.
It constitutes a good description without considering more complex effects, such as the
contributions from coherent scattering from pairs
of impurity centers, which requires a quantum transport theory~\cite{moore1967quantum}.

The per-unit-time transition probability for the scattering of charge carriers
by ionized impurities can be written in the plane-wave approximation as 
\begin{multline}
\label{U}
W({\bf{k}}|{\bf{{k}^{\prime}}}) = \frac{2\pi}{\hbar}\frac{N_i}{V} \\
\left\vert{\int U({\bf{r}}) \exp\left[i({\bf{k}}-{\bf{{k}^{\prime}}})\cdot{\bf{r}}\right]d{\bf{r}}}\right\vert^2 \delta(\epsilon_{{\bf{{k}^{\prime}}}}-\epsilon_{{\bf{k}}})~,
\end{multline}
where $U({\bf{r}})$ is the scattering potential and $N_i$ is the ionized impurity concentration.

The long-range Coulomb field,
$U(\mat{r}) = e\phi(r) = \pm e^2/\zeta{_0} r$, where the potential $\phi$ at a point $r$ of the crystal 
is due to the presence of positive (donor) or negative (acceptor)
impurity ions. The straightforward application of this field in Eq.~\eqref{U} leads
to a logarithmic divergence, and hence, a screened Coulomb potential has to be considered.
According to the B-H theory, the potential can be expressed in a more rigorous form as 
$\phi(r) = \pm e/{\zeta{_0} r} \left(\exp\left(-{r}/{r_0}\right)\right)$, 
where $r_0$ is the radius of ion field screening defined by Eq.~\eqref{r0}.
From Eqs.~(\ref{tau3}),~(\ref{1PT}), and~(\ref{U}), the RT for the scattering of
charge carriers by ionized impurities can be expressed for each band $n$ as 
\begin{equation}
\label{tau_imp}
\tau_{imp}(k) = \frac{\hbar\zeta{_0}{^2}}{2{\pi}{e^4}{N_i}F_{imp}(k)}k^2 \left\vert\frac{\partial \epsilon_k}{\partial k}\right\vert 
\end{equation}
where
\begin{equation}
F_{imp}(k) = \ln(1+\eta) - \frac{\eta}{1+\eta}~, 
\end{equation}
is the screening function, with $\eta = (2kr_0)^2$.

\section{Computational Implementation}

The evaluation of carrier transport properties
were carried out using the \texttt{BoltzTraP} code as a reference~\cite{madsen2006boltztrap}.
The \texttt{BoltzTraP} code solves the linearized BTE by Fourier interpolating
the band structure computed within the DFT framework 
and performs all the integrations required to determine the TE transport properties. 

\subsubsection{Fourier Interpolation}

In order to introduce the main ideas behind Fourier interpolation 
within \texttt{BoltzTraP}, let us consider $N_{KS}$ KS eigenvalues for a 
given band $n$ 
of a three-dimensional (3D) periodic solid.
The symmetry of the crystal's reciprocal space is incorporated in the energy bands. 
Therefore, it is natural to use star functions, $\Upsilon_m({\bf{k}})$, 
as basis set to Fourier expand the quasi-particles energies 
\begin{equation}
\tilde{\epsilon}_{{\bf{k}}}=\sum_{m=1}^M a_m\Upsilon_m({\bf{k}})~,
\end{equation}
where 
\begin{equation}
\Upsilon_{m}({\bf{k}})=\frac{1}{n_s}\sum_{\{\upsilon\}}\exp[i(\upsilon {\bf{R}}_m)\cdot{\bf{k}}]~,
\end{equation}
with the sum running over all $n_s$ point group operations $\{\upsilon\}$ 
on the direct lattice translations, ${\bf{R}}_m$.
The first derivatives are straightforwardly given by
\begin{equation}
\label{velocities}
{\bf{v}}_{{\bf{k}}}=\frac{\partial {\tilde{\epsilon}_{{\bf{k}}}}}{\partial {\bf{k}}} = \frac{i}{n_s}\sum_{m=1}^M a_m \sum_{\{\upsilon\}}(\upsilon {\bf{R}}_m) \exp [i(\upsilon {\bf{R}}_m)\cdot{\bf{k}}]~,
\end{equation}
in which the main problem is the determination of Fourier coefficients, $a_m$.

\texttt{BoltzTraP} relies on the proposal by Shankland~\cite{shankland1971interpolation},
according to which one should choose a set of basis functions for interpolation
larger than the number of data points ($M>N_{KS}$) and constrain
the interpolation function to pass exactly through such points.
In order to obtain a smooth interpolation,
the extra basis functions are used to minimize
a roughness function suitably defined by Pickett, Krakauer and Allen~\cite{pickett1988smooth} 
\begin{equation}
\label{R}
\mat{\Re} = \sum_{m=2}^M \lvert a_m \lvert ^2 \rho(R_m)
\end{equation}
with
\begin{equation}
\label{roughness}
\rho(R_m) = \left(1-c_1\left(\frac{R_m}{R_{min}}\right)^2\right)^2+c_2\left(\frac{R_m}{R_{min}}\right)^6~,
\end{equation}
where $R_m = \lvert {\bf{R}}_m \lvert$, $R_{min}$ is the magnitude 
of the smallest nonzero lattice vector, and $c_1 = c_2 = 3/4$.
Thus, the Lagrange multiplier method can now be used
since the formulated problem is to
minimize $\mat{\Re}$ subject to the constraints, 
${\tilde{\epsilon}}_{{\bf{k}}_l} = \epsilon_{{\bf{k}}_l}$, with respect to the Fourier coefficients. 
From such minimization one obtains
\begin{equation}
a_m = \begin{cases}
\rho(R_m)^{-1} \sum_{l=1}^{N_{KS}-1}\lambda^*_l\left[\Upsilon_m^{*}({\bf{k}}_l) - \Upsilon_m^{*}({\bf{k}}_{N_{KS}})\right],  &  m>1, \\

\epsilon_{{\bf{k}}_{N_{KS}}}-\sum_{m=2}^M a_m \Upsilon_m ({\bf{k}}_{N_{KS}}),  &  m=1,
\end{cases}
\end{equation}
in which the Lagrange multipliers, $\lambda^*_l$, can be evaluated from 
\begin{equation}
\epsilon_{{\bf{k}}_p} - \epsilon_{{\bf{k}}_{N_{KS}}} = \sum_{l=1}^{N_{KS}-1}{\bf{H}}_{pl}\lambda^*_l~,
\end{equation}
with 
\begin{equation}
{\bf{H}}_{pl} = \sum_{m=2}^M \frac{\left[\Upsilon_m({\bf{k}}_p) - \Upsilon_m({\bf{k}}_{N_{KS}})\right]\left[\Upsilon_m^{*}({\bf{k}}_l)-\Upsilon_m^{*}({\bf{k}}_{N_{KS}})\right]}{\rho(R_m)}~.
\end{equation}

\subsubsection{Anisotropic ${\bf{k}}$-mesh}

In practice, a finer and anisotropic ${\bf{k}}$-mesh can be generated for the interpolation. 
To acomplish this, lattice points and their respective star functions 
are generated in real space 
following point the group operations of the crystal symmetry. The 
corresponding translation vector can be given as  
${\bf{R}} = u_1{\bf{a_1}} + u_2{\bf{a_2}} + u_3{\bf{a_3}}$, in which 
${\bf{a_1}}$, ${\bf{a_2}}$, ${\bf{a_3}}$ are related to the crystal primitive vectors. 
Such points are generated inside a sphere of a radius 
$R^{\prime} = \sqrt[3]{3\cdot n_{kpt}\cdot (M/N_{KS}) \cdot \Omega/4\pi}$,   
where n$_{kpt}$ is the number of ${\bf{k}}$-points of 
the original ${\bf{k}}$-mesh in the entire BZ, $(M/N_{KS})$
is the required number of star functions per ${\bf{k}}$-point 
and $\Omega$ is the volume of the unit cell.
Consequently, $R^{\prime}$ determines the full extension of the real space and can be 
properly changed, for example, by increasing the number of star functions per ${\bf{k}}$-point. 
The corresponding reciprocal lattice, with a translation vector, 
${\bf{k}} = k_1{\bf{b_1}} + k_2{\bf{b_2}} + k_3{\bf{b_3}}$ can be determined 
by generating its three primitive vectors from the direct ones from 
$\left[{\bf{b_1}} {\bf{b_2}} {\bf{b_3}}\right]^T = \left[{\bf{a_1}} {\bf{a_2}} {\bf{a_3}}\right]^{-1}$,
in which the $2\pi$ factor was omitted following the crystallographic 
definition of reciprocal space. 
In order to capture the crystal anisotropy, 
the extension of the real space can be determined for each 
crystal direction, defining spheres for each crystallographic axis with the maximum radius given by 
$R_{max}(t) = INT(R^{\prime}\cdot\sqrt{{\bf{b_t}}\cdot{\bf{b_t}}}) + 1$, 
with $t=\{1,2,3\}$ and $INT(x)$ gives the largest integer number 
that does not exceed the magnitude of $x$. 
Hence, lattice points fill all the spheres space inside the $\{-R_{max}(t),R_{max}(t)\}$ range. 
Within \texttt{BoltzTraP}, a 3D array containing all vectors 
are sorted considering their 
concentric radius, $r$, from the sphere center, and providing that 
all vectors, ${\bf{R}}$, have different star functions, $m$. In practice, 
to determine all $\bf{k}$ vectors from $\bf{R}_m$, 
a 3D Fast Fourier Transform (FFT) is performed, 
being possible to redefine a finer and anisotropic $\bf{k}$-mesh. 

The magnitude of each $\bf{R}_m$ vector, $r$, is defined 
through the metric tensor formalism.
Let us consider the scalar product of two arbitrary vectors
in the coordinate system of the real space,
${\bf{r_1}}\cdot{\bf{r_2}} = (x_1{\bf{a_1}}+y_1{\bf{a_2}}+z_1{\bf{a_3}})\cdot(x_2{\bf{a_1}}+y_2{\bf{a_2}}+z_2{\bf{a_3}})$.
In the matrix notation this is written as
\begin{multline}
{\bf{r_1}}\cdot{\bf{r_2}}=
\begin{bmatrix} 
    x_{1} & y_{1} & z_{1}
\end{bmatrix}
\begin{bmatrix}
    {\bf{a_1}}\cdot{\bf{a_1}} & {\bf{a_1}}\cdot{\bf{a_2}} & {\bf{a_1}}\cdot{\bf{a_3}} \\
    {\bf{a_2}}\cdot{\bf{a_1}} & {\bf{a_2}}\cdot{\bf{a_2}} & {\bf{a_2}}\cdot{\bf{a_3}} \\
    {\bf{a_3}}\cdot{\bf{a_1}} & {\bf{a_3}}\cdot{\bf{a_2}} & {\bf{a_3}}\cdot{\bf{a_3}}
\end{bmatrix}
\begin{bmatrix} 
    x_{2} \\
    y_{2} \\
    z_{2}
\end{bmatrix}
\\
={\bf{\bar{X}_1}}\bf{G}\bf{{X}_2}.
\end{multline}
Considering that ${\bf{r_1}} = {\bf{r_2}} = {\bf{R}}_m$, the magnitude of the
real space vector is given by 
$r=\sqrt{\bf{\bar{X}}\bf{G}\bf{X}}$, 
with ${\bf{\bar{X}}} = [u_1 u_2 u_3]$.
For all $\bf{R}_m$ in the Bravais lattice, the reciprocal lattice is characterized 
by a set of wavevectors ${\bf{k}}$, such that, $e^{2\pi i {\bf{k}}\cdot{\bf{R}}_m} = 1$. 
Given ${\bf{R}}_m$ and ${\bf{k}}$ in the same direction, the 
magnitude of the vector ${\bf{k}}$ in the reciprocal space is given by
$\left\vert {\bf{k}} \right\vert = (k_1u_1 + k_2u_2 + k_3u_3)/r = n_{int}(1)+n_{int}(2)+n_{int}(3)/r$, 
where $n_{int}(t) = 1,2,...,k_{max}(t)$ 
are integer numbers with $k_{max}(t)=2R_{max}(t)+1$ defining the $\bf{k}$-mesh grid. 
In practice, $k_{max}(t)$ should be carefully taken as the product of small primes in order 
to improve the efficiency of FFT. 
If we consider just one specific crystallographic axis,
the magnitude of the $\bf{k}$ vector
can be determined for each axis by $\left\vert{{\bf{k}}}\right\vert_t = k_t = n_{int}(t)/r$. 
Hence, through the definition of anisotropic ${\bf{k}}$-mesh
the anisotropy of materials TE properties can be captured.

\subsubsection{Anisotropic RTs}

The consideration of anisotropy in the RTs of the aforementioned scattering processes
can now be introduced.
We start by generalizing the closed expressions for all RTs formulae, 
comprising Eqs.~(\ref{tau_ac}),~(\ref{tau_npol}),~(\ref{tau_pol}),~(\ref{tau_imp}),  
for every band $n$ by considering the interpolated quasi-particles 
energies, $\tilde\epsilon_{n,{\bf{k}}}$, and their derivatives. 
In such expressions for the RT, the explicit dependence on ${\bf{k}}$ can have an
arbitrary form, beyond the parabolic dependence.
Hence, we may properly account for the multiplicity, degeneracy and 
non-parabolicity of the band edges on the same footing, along with anisotropy.  
The expression for the RT of the non-polar scattering by acoustic phonons 
can now be expressed along the direction $t$ and for a given band index $n$
\begin{equation}
\label{tau_ac2}
(\tau_{ac})_t(n,k) = \frac{\pi\hbar\rho}{(E{_{1})_t}^2}\frac{(v_{0})_t^2}{k{_B}T}\frac{1}{k_t^2}\left\vert\frac{\partial \epsilon_{n,k}}{\partial k_t}\right\vert~,
\end{equation}
in which the parameters have to be taken 
along each crystallographic axis, i.e., it is necessary to consider 
anisotropic parameters. Hence, the subscript $t$ in Eq.\eqref{tau_ac2}, as well
in the subsequent equations, indicates a given crystallographic direction. 
The acoustic deformation potential for holes and electrons can be calculated 
from \textit{ab initio} methods on the basis of the energy change of the 
VBM and CBM, respectively, with respect to a deep 
core state that is practically insensitive to slight lattice deformations.  
The anisotropic velocity entering in Eq.(\eqref{tau_ac2}) as 
well as in the subsequent equations is defined as
\begin{multline}
\label{velocities2}
v_{t}(n,{\bf{k}}) = \left(\frac{\partial {\epsilon}_{n,{\bf{k}}}}{\partial {\bf{k}}}\right)_t 
\approx \left(\frac{\partial {\tilde{\epsilon}_{n,{\bf{k}}}}}{\partial {\bf{k}}}\right)_t = \\
\frac{i}{n_s}\sum_{m=1}^M a_m \sum_{\{\upsilon\}}(\upsilon {\bf{R}}_{m})_t \exp [i(\upsilon {\bf{R}}_{m})\cdot{\bf{k}}]~, 
\end{multline}
in which we are only interested in the magnitude of the velocity.  

Along the same line of reasoning, in order to capture the anisotropy, 
the RT of the non-polar scattering 
by optical phonons can now be rewritten as
\begin{equation}
\label{Tau_npol2}
(\tau_{npol})_t(n,k) = \frac{\beta_t^2}{\pi\hbar}\frac{{\rho}a_t^2}{k{_B}T}\frac{1}{k_t^2}\left\vert\frac{\partial \epsilon_{n,k}}{\partial k_t}\right\vert~.  
\end{equation}
The evaluation of the anisotropic ODP from \textit{ab initio} methods is more involving 
than the evaluation of the ADP, since it characterizes the energy change of the carrier 
due to a relative atomic displacement. Therefore, 
external stresses that deform the unit cell as a
whole cannot reveal ODPs, which have to be evaluated empirically 
on the basis of experimental data. 
The anisotropic RT of polar mode scattering of 
optical phonons can also be written as
\begin{equation}
\label{Tau_pol}
(\tau_{pol})_t(n,k) = \frac{K_{pol}F_{pol}\zeta_t\hbar}{2{e^2}k{_B}T}\left\vert\frac{\partial \epsilon_{n,k}}{\partial k_t}\right\vert~,  
\end{equation}
with the anisotropic screening function given by 
\begin{multline}
\label{screen_pol2}
F_{pol} = \\
\left[1 -\frac{1}{2(r_{0,t}{k_t})^2}\ln[1+4(r_{0,t}{k_t})^2]+\frac{1}{(1+4(r_{0,t}{k_t})^2}\right]^{-1}~,
\end{multline}
in which $r_{0,t}$ is the anisotropic screening radius defined from Eq.\eqref{r0} as
\begin{equation}
\label{r02}
r_{0,t}^{-2}(n,k) = \frac{4\pi{e^2}}{\zeta_{0,t}}\int{-\frac{\partial f_0(\epsilon,\mu_t,T)}{\partial \epsilon_{n,k}}}g(\epsilon)d\epsilon~, 
\end{equation}
where it has been included the dependence on the band index $n$. The equilibrium electron
distribution function, $f^{(0)}(\epsilon,\mu_t,T)$, is also defined to be 
anisotropic through an anisotropic electronic 
chemical potential $\mu_t$, which will be discussed in detail below. In practice, the 
evaluation of density of states, $g(\epsilon)$, 
is obtained numerically on an energy grid 
with spacing $d\epsilon$ sampled over $N_k$ ${\bf{k}}$-points
\begin{equation}
\label{dos2}
g(\epsilon) = \int \sum_n \delta(\epsilon - \epsilon_{n,{\bf{k}}}) \frac{d{\bf{k}}}{8\pi^3} = \frac{1}{\Omega N_k}\sum_{n,{\bf{k}}} \frac{\delta(\epsilon - \epsilon_{n,{\bf{k}}})}{d\epsilon}~.
\end{equation}
Similarly, the expression for the RT of scattering by ionized impurities 
has to be rewritten to include anisotropy as
\begin{equation}
\label{Tau_imp2}
(\tau_{imp})_t(n,k) = \frac{\hbar\zeta_{0,t}{^2}}{2{\pi}{e^4}{N_{i,t}F_{imp}(n,k_t)}} {k_t}^2 \left\vert\left(\frac{\partial \epsilon_{n,k}}{\partial k_t}\right)\right\vert~,
\end{equation}
where the screening function has to be rewritten as 
$F_{imp}(n,k_t) = \ln(1+\eta) - {\eta}/({1+\eta})$, with $\eta=(2{k_t} r_{0,t})^2$, 
in which the anisotropic screening radius, $r_{0,t}$, has been defined by Eq.\eqref{r02}.  

The electronic chemical potential, $\mu$, 
is a result of the condition of charge neutrality.
In principle, $\mu$ can be experimentally influenced mainly by doping, 
which determines the carrier concentration. 
The anisotropic electronic chemical potential, $\mu_t$ , as entering in the expressions for the
RTs, $(\tau_{pol})_t$ and $(\tau_{imp})_t$ , through their respective screening formulae, is a direct consequence of dissimilar 
carrier concentrations along different axial directions, as it has been measured by Zhao et al.~\cite{zhao2014ultralow} 
through Hall transport properties.
It is important to stress that such measurements have been done along each axial direction separately and it does not mean three 
different chemical potentials simultaneously in a single system. The same line of reasoning has been adopted here by defining 
an anisotropic chemical potential, given that our calculations are performed independently for each axial direction.
Given $\mu_t$ and T, the carrier concentration in the axis $t$
can be calculated as the deviation from charge neutrality
\begin{equation}
\label{charge_neutrality}
n_{carr}(\mu_t,T) = Z - \int g(\epsilon)f^{(0)}(\epsilon;\mu_t;T)d\epsilon~,
\end{equation}
in which $Z$ is the nuclear charge and $g(\epsilon)$
is the density of states as evaluated from Eq.~\eqref{dos2}. 
In practice, we can insert a target carrier concentration $n_{carr}^{(0)}$ within \texttt{BoltzTraP},
with $n_{carr}^{(0)} > 0$ ($n_{carr}^{(0)} < 0$) for holes (electrons)
and $\mu_t$ is calculated iteratively up to $n_{carr}(\mu_t,T) \sim n_{carr}^{(0)}$ 
within a predefined criterium of $10^{-12}$.
The incremental variation of $\mu_t$ is given by
\begin{equation}
\label{increment}
\Delta\mu_t = \frac{n_{carr}(\mu_t,T) - n_{carr}^{(0)}}{\int g(\epsilon) \frac{df^{(0)}}{d\epsilon}d\epsilon}~,
\end{equation}
and hence, a $n$-type semiconductor can be obtained
if $\mu_t$ approaches the conduction band while
a $p$-type material is obtained when $\mu_t$ moves to
the valence band. 

Once obtained the anisotropic RTs for each scattering process, 
the total anisotropic RT, $(\tau_{tot})_t$, that enters in the TE transport calculations 
can be derived from the Mathiessen's rule as  
\begin{multline}
\label{Mathiessen}
(\tau_{tot})_t^{-1}(n,k,\mu_t,T) = (\tau_{ac})_t^{-1}(n,k,T) + (\tau_{npol})_t^{-1}(n,k,T) \\
+ (\tau_{pol})_t^{-1}(n,k,\mu_t,T) + (\tau_{imp})_t^{-1}(n,k,\mu_t,T)~,
\end{multline}
which is justified by considering the scattering mechanisms as approximately independent.
Once having $(\tau_{tot})_t$, the TDKs can be evaluated
similarly to the calculation of the DOS as 
\begin{multline}
\Xi_{tt}(\epsilon,\mu_t,T) = \frac{e^2}{\Omega N_k} \sum_{n,{\bf{k}}} (\tau_{tot})_t(n,k,\mu_t,T) \\
{\bf{v}}_{n,{\bf{k}}}{\bf{v}}_{n,{\bf{k}}}^T \frac{\delta(\epsilon - \epsilon_{n,{\bf{k}}})}{d\epsilon}~, 
\end{multline}
where ${\bf{v}}_{n,{\bf{k}}}^T$ is the transpose of the vector ${\bf{v}}_{n,{\bf{k}}}$ given by Eq.(\ref{velocities}) 
expanded over the band indices, 
and the matrix ${\bf{v}}_{n,{\bf{k}}}{\bf{v}}_{n,{\bf{k}}}^T$ is diagonal by construction. 
From the TDKs, the coefficient transport tensors, given by Eq.\eqref{Lambda}, can then be evaluated. 
For example, $\Lambda_{tt}^{(0)}(\mu_t,T) = \sigma_{tt}(\mu_t,T)$ is given by 
\begin{equation}
\sigma_{tt}(\mu_t,T)= \int \Xi_{tt}(\epsilon,\mu_t,T) \left[\frac{-\partial f^{(0)}(\epsilon,\mu_t,T)}{\partial\epsilon}\right]d\epsilon~,
\end{equation}
which accounts for the diagonal terms of the electrical conductivity tensor. 
The diagonal terms of Seebeck coefficient tensor ($S_{tt}(\mu_t,T)$) is obtained by 
calculating $\Lambda_{tt}^{(1)}(\mu_t,T)$ as 
\begin{multline}
\Lambda_{tt}^{(1)}(\mu_t,T)= \int \Xi_{tt}(\epsilon,\mu_t,T) (\epsilon - \mu_t)\\
\left[\frac{-\partial f^{(0)}(\epsilon,\mu_t,T)}{\partial\epsilon}\right]d\epsilon~,
\label{Seebeck}
\end{multline}
in which $S_{tt}(\mu_t,T) = (eT)^{-1}\Lambda_{tt}^{(1)}(\mu_t,T)\cdot[\sigma_{tt}(\mu_t,T)]^{-1}$. 
Similarly, the diagonal terms of the corresponding coefficient tensor for the charge carrier 
contribution to the thermal conductivity, $\kappa_{elec}$, can also be determined. 
Its is important to note that, due to the factor 
$\left[-\partial f^{(0)}/\partial\epsilon\right]$, only bands within
a few $k_BT$ from the Fermi level are relevant for transport properties. 
 
\section{Results: The Case of Anisotropic \ce{SnSe}}

The methodology developed in the preceding sections was employed
to investigate TE transport properties 
of p-type tin selenide (SnSe), which is  
one of the highest-performing TE materials available today. 
To perform the calculations, we have considered the RT models 
for scattering mechanisms introduced earlier. 
The polar scattering by acoustic phonons, 
known as piezoelectric scattering, has been neglected here. 
As emphasized by Li~\cite{li2012semiconductor}, for a typical III-V compound semiconductor,
the piezoelectric scattering is usually much less important than the acoustic
deformation potential scattering. The piezoelectric scattering becomes more
important with increasing ionicity of the systems, for example,
in most of the II-VI compound semiconductors, in which the wurtzite crystal
structure lacks inversion symmetry, and, therefore, the piezoelectric stress tensor 
is nonvanishing. On the basis of the facts above, it is expected that
this kind of scattering mechanism is not relevant for IV-VI SnSe.\\

The choice to  study SnSe  
is due to its following features. 
$i)$ The charge carrier transport properties of \ce{SnSe}
have a strong anisotropy, which is expected due to its layered structure.
$ii)$ In particular, p-type \ce{SnSe} presents strongly
non-parabolic dispersion and its VBM has a \textit{pudding-mold-like}
shape~\cite{kutorasinski2015electronic}.
$iii)$ It is also noticed that the first and second VBMs can be considered degenerate,
with an offset between them of only \num{0.02} - \num{0.06} eV~\cite{zhao2015ultrahigh}.
$iv)$ \ce{SnSe} exhibits multiple local band edges aside from the global extrema,
within an energy range comparable to the thermal energy of the carriers~\cite{shi2015quasiparticle}.
$v)$ Moreover, SnSe has attracted much attention in recent years due to its record $zT$
$\sim\num{2.6}$~\cite{zhao2014ultralow}.
Therefore, based on the above features, we found \ce{SnSe}
to be a suitable material to test the methodology we are proposing here. Our 
results are compared with the experimental data
provided by Zhao \textit{et al.}~\cite{zhao2014ultralow,zhao2015ultrahigh}.

At room temperature, SnSe crystallizes in a layered orthorhombic structure
with the \textit{Pnma} space group and $\num{8}$ atoms in the unit cell.
The nomenclature of the SnSe axes will follow 
that employed in the work of Zhao \textit{et al.}
Along the $b$-$c$ plane, Sn and Se atoms are linked by strong covalent bonds ($\sim\num{2.80}\:\AA$)
forming a layered substructure with zig-zag chains along the $b$ axis.
The intralayer bonds are much stronger than the interlayer interactions.
The latter are predominantly weak van der Waals-like
Se-Sn bonds ($\sim\num{3.50}\:\AA$) along the $a$-axis.
Upon heating, SnSe undergoes a
second-order phase transition to the higher symmetry \textit{Cmcm}
phase around T$\sim$$\num{807}$~K,
due to the condensation of the TO soft
phonon mode of $A_{g}$ symmetry at the BZ center ($\Gamma$).
However, in this work, we are considering only the low-temperature \textit{Pnma} phase of SnSe, and, therefore,
from now on, unless otherwise explicitly stated, SnSe will be referred only to
this phase, without reference to the high-temperature phase.

\subsection{DFT Calculation of Crystalline Structure and Band Structure}

For the optimization of the crystalline structure and band structure calculations, we employed
DFT within the generalized gradient approximation (GGA) 
in the formulation of Perdew-Burke-Ernzerhof 
(PBE)~\cite{perdew1996generalized}, as implemented in the VASP package~\cite{kresse1993ab,kresse1996efficient}.
To model the ion cores, projector augmented-wave (PAW)
pseudopotentials~\cite{kresse1999ultrasoft} were used, considering as valence electrons
those in the orbitals $4d$, $5s$, and $5p$ of Sn and those in the orbitals $4s$ and $4p$ of Se.
The electronic wave functions were expanded in a plane-wave basis-set
with a kinetic-energy cutoff of \num{800} eV. To sample the BZ,
a 4 x 11 x 12 Monkhorst-Pack ${\bf{k}}$-point 
mesh was used for the force on the ions calculations. 
Subsequently, to determine the transport properties, we used a finer ${\bf{k}}$-mesh 
of 16 x 40 x 43, which ensures the calculated transport coefficients to be converged. 
Within \texttt{BoltzTraP} the original ${\bf{k}}$-mesh is then interpolated
onto a mesh five times as dense ($M/N_{KS} = 5$).
We used experimental structure~\cite{adouby1998structure} as starting configuration and
relaxed the lattice parameters and atomic positions until
all atomic force components were smaller than $\num{1}$ meV/$\AA$.

The weak interactions between layers along $a$ direction were
accurately captured by incorporating van der Waals (vdW) corrections to DFT, 
according to the D3 approach as provided by Grimme \textit{et al.}~\cite{grimme2010consistent}.
We found that DFT-D3 performed reasonably in predicting both the lattice
parameters and the internal atomic positions of SnSe. 
We found that $a = 11.564\:\AA$, $b = 4.542\:\AA$ and $c = 4.166\:\AA$, 
diverging by $0.52$\%, $1.98$\% and $0.24$\%, respectively, from experimental lattice 
structure parameters determined at 300K~\cite{adouby1998structure}. 
The spin-orbit coupling (SOC) has been neglected in our calculations. 
As shown by Wu \textit{et al.}~\cite{wu2017engineering} the neglect SOC
does not lead to noticeable effects on the band structures, which has been attributed to
the low symmetry of SnSe that removes the degeneracy of the electronic states.

The optimization of TE properties by chemical doping has been widely adopted.
Particularly for SnSe, hole-doping by p-type dopants
such as Ag~\cite{chen2014thermoelectric,leng2016optimization,peng2016broad}
and Na~\cite{peng2016broad}
have led to enhancement of $zT$ over a broad
temperature range in comparison with the intrinsic (undoped) material.
Here, we have also considered transport properties of
hole-doped SnSe. An important modification caused in SnSe due to doping 
is the downward shift of chemical potential that enters in the 
valence band, turning the crystal into a metal.
However, we have considered that the valence band shape has not changed considerably 
in relation to the intrinsic case due to doping,
the so-called rigid band approximation. 

\subsection{Calculation of SnSe TE Transport Properties}

\subsubsection{Carrier Concentration for Intrinsic and Extrinsic P-Type Cases}

\begin{figure}
	\centering
	\includegraphics[width=0.45\textwidth]{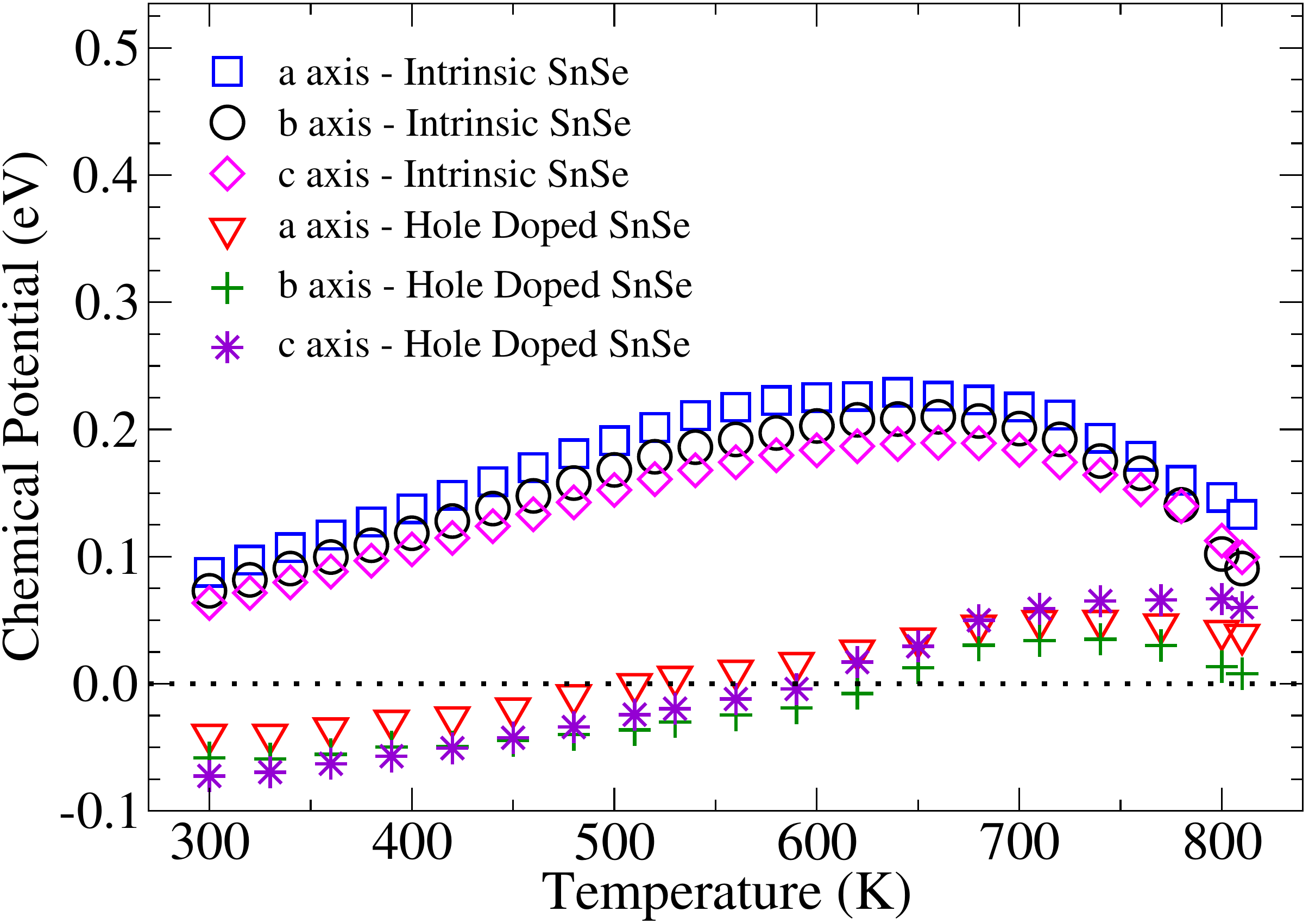}
	\caption{Evolution of the chemical potential with temperature ($\num{300}$-$\num{810}$~K) 
                 for both intrinsic and hole-doped \ce{SnSe} 
		 for each crystallographic axis.}
	\label{EF}
\end{figure}

In order to calculate TE transport properties for p-type SnSe, firstly it is
necessary to properly estimate the hole concentration.
On the basis of experimental anisotropic \ce{SnSe} Hall coefficients~\cite{zhao2014ultralow,zhao2015ultrahigh},
from which Hall concentration, $n_{Hall}$, can be inferred for each axis,
it is possible to acquire hints concerning the evolution
of hole concentration with temperature.
In particular, for intrinsic \ce{SnSe}, Hall concentration 
stays almost constant within the range of $\sim\num{300}-\num{550}$~K, 
corresponding to non-equilibrium concentration of charge carriers due to 
growth conditions. At higher temperatures comprising the transition 
region ($\sim\num{550}-\num{807}$~K), a marked Arrhenius-type 
enhancement has been observed that is mainly related to 
thermally activated formation of vacancies accompanied by the production of holes~\cite{dewandre2016two}. 
The formation energies of intrinsic defects in SnSe has been systematically
investigated on the basis of DFT~\cite{dewandre2016two,huang2017first}.
The Sn vacancy, V$_{Sn}$, exhibited the lowest formation energy
with an ultra-shallow thermodynamic transition level. Hence, V$_{Sn}$
plays a prominent role for p-type conduction in the absence of extrinsic doping. 
For example, depending upon the chemical potential, the \ce{Sn} vacancy can be singly, V$_{Sn}^{(1-)}$, or 
doubly, V$_{Sn}^{(2-)}$, charged,
by capturing one or two electrons and producing one or two holes for each created vacancy~\cite{dewandre2016two}.\\
\begin{figure}
	\centering
	\includegraphics[width=0.45\textwidth,viewport= 0 125 600 890,clip]{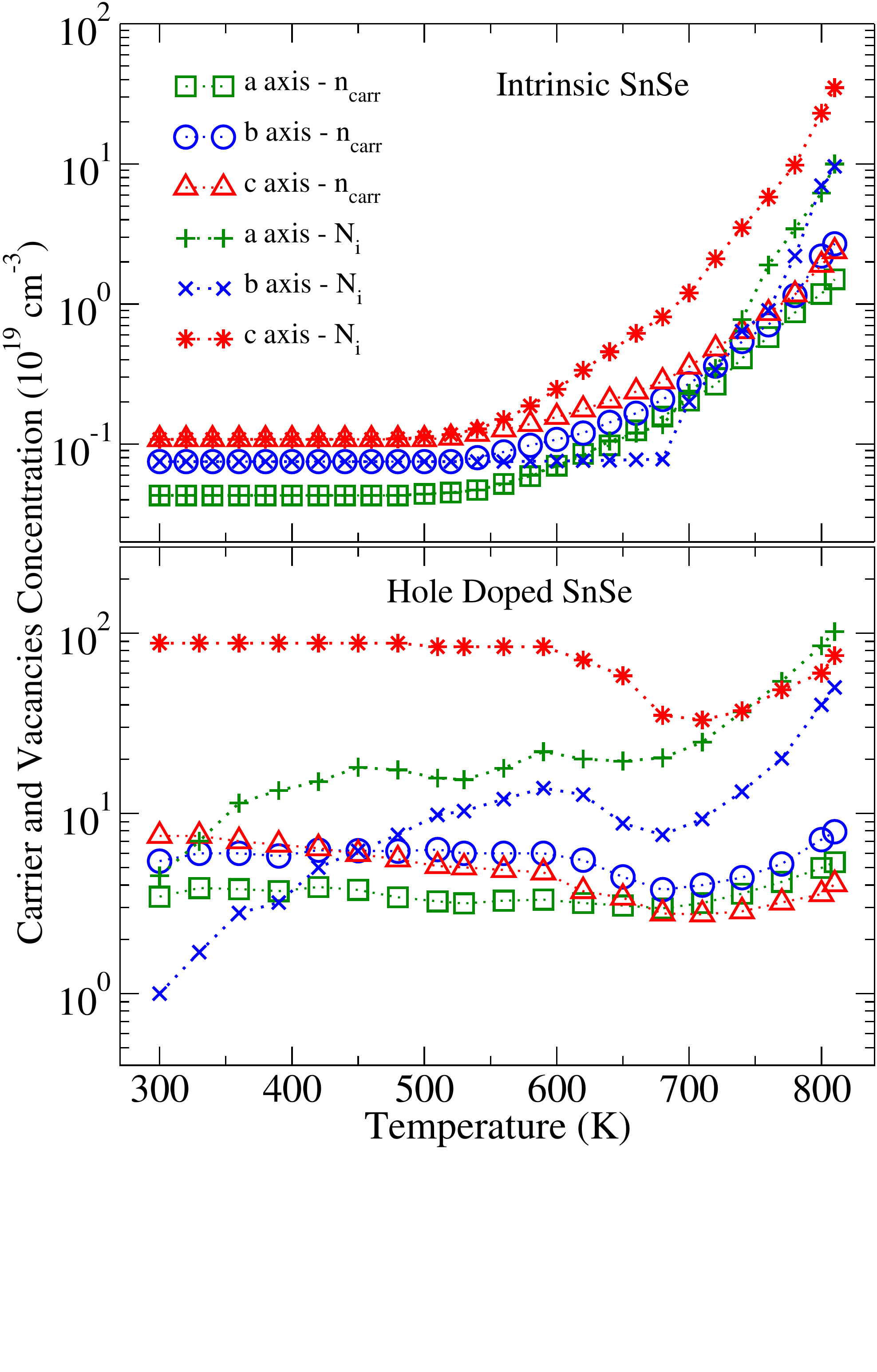}
	\caption{Evolution of carrier ($n_{carr}$) and vacancies ($N_i$)
		concentrations with temperature ($\num{300}$-$\num{810}$~K) in intrinsic \ce{SnSe} (top panel)
		and in hole-doped \ce{SnSe} (bottom panel) for each crystallographic axis.}
	\label{conc}
\end{figure} 
\indent As pointed out by Kutorasinski \textit{et al}~\cite{kutorasinski2015electronic}, Hall concentrations
should not be considered the actual hole concentrations because 
it is based on the assumption of a single parabolic band model and Hall scattering factor of unity.
Hence, these measurements yield only an indication of the charge carrier concentrations.
Despite of this qualitative aspect, the evolution of the
experimental Hall coefficients suggests that the carrier
concentration increases with temperature.
Here, in order to determine the hole concentration, $n_{carr}$, 
we applied the condition that experimental Seebeck coefficients (S$_{exp}$)~\cite{zhao2014ultralow,zhao2015ultrahigh}
should be essentially the same as theoretical ones (S$_{theor}(\mu,T)$), that is,
S$_{exp} =$ S$_{theor}(\mu,T)$ for any temperature. This fitting is possible 
by varying the target hole concentration in the calculation 
of the Seebeck coefficients in Eq.\eqref{Seebeck}. This is done 
through the neutrality condition (Eq.\eqref{charge_neutrality}); for a given concentration $n_{carr}(\mu_t,T)$ 
a corresponding chemical potential, $\mu_t$, for 
each axis, is obtained. As $n_{Hall}$ suggests that $n_{carr}$ 
should be almost constant in the range of $\sim\num{300}-\num{550}$~K, we kept 
constant $n_{carr}$ in this range, which best adjust the experimental data. 
The resulting hole concentration, $n_{carr}$, is shown in the top panel of Fig.~\ref{conc}.\\
\indent The same methodology has been applied for the case of 
hole-doped \ce{SnSe}, in which the resulting hole concentration is 
shown in the bottom panel of Fig.~\ref{conc}. In this case, the 
evolution of hole concentration is much more involved and
includes complex processes, such as the compensation effects, possibly
due to the formation of doubly charged \ce{Se} vacancies, V$_{Se}^{(2+)}$ and antisite defects, Se$_{Sn}^{(2+)}$, which will annihilate
the holes created by acceptor impurities, such as Na, and vacancies,
or due to the change in the charge state of the impurities.
In fact, the measured Hall coefficients for the $b$ and $c$ axes highlights the tendency of
the hole concentration to decrease with temperature, especially around \num{600}~K 
Hall coefficient abruptly raises~\cite{zhao2015ultrahigh}, in close correspondence with our results.  
Furthermore, the same tendency was also observed
in \ce{Na} and \ce{Ag}-doped p-type \ce{SnSe}, an specific kind of hole-doped \ce{SnSe}~\cite{Peng}.
Unlike metals, in which the conductivity decreases almost inversely with temperature,
the conductivity of hole-doped SnSe decreases with the temperature following an exponential-like decay,
suggesting that the carrier concentration are diminishing, 
as shown in the lower panel Fig.~\ref{conc}.
Withal, the chemical potential shifts up with temperature as indicated by Fig.~\ref{EF}.
Eventually, at elevated temperatures, an activated
(semiconducting) behavior sets in, leading $n_{carr}$ to increase, 
mainly above \num{700}~K.\\
\indent In the intrinsic case, at low temperatures ($T\lesssim550$~K), the
concentration of ionized impurities, $N_i$, is due to crystal growth 
conditions corresponding to a non-equilibrium concentration of vacancies. 
In this temperature range $N_i$ has been considered to be almost 
the same as $n_{carr}$, since according to previous calculations~\cite{dewandre2016two}, in the range of 
chemical potential determined in our calculations, the Sn vacancy is predominantly singly negatively charged, V$_{Sn}^{(1-)}$.
However, at a given higher temperature, the
thermally activated process of vacancy formation becomes relevant and we have to 
take into account the variation of $N_i$ with temperature. This is done in our calculations by adjusting our calculated electrical conductivity
to experimental results. In other words, the optimal values of $N_i$ presented in the top panel Fig.~\ref{conc},
are the result of this fitting procedure.\\
\indent The concentrations of ionized impurities for the extrinsic hole-doped SnSe are shown in the bottom panel of Fig.~\ref{conc}.
As it was mentioned before, in the extrinsic case there are different charged impurities at play and the overall behavior is complex.
The concentrations for the crystalline axes are somewhat related to their chemical potentials. It is important to notice that as temperature rises and
the chemical potential becomes positive (enters the gap), there is a dip in the concentration of ionized impurities, this reduction 
is possibly caused by a change in the charge state of the impurities, which become neutral. Another interesting feature is that as temperature
continues rising, the concentrations of ionized impurities increase steeply, possibly due to the creation of charged impurities, analogously 
to the intrinsic case, and the concentrations for each one of the axes seem to converge to a similar value.

\subsubsection{SnSe Parameters for the RT Models}


Once the temperature evolution of hole concentration 
is determined, it is possible to 
adjust RT models for TE transport modeling of SnSe. 
The parameters entering in the RT models for both cases of intrinsic and hole-doped SnSe  
are shown in Table~\ref{parameters}. In particular, ADP constants ($E_1$) have been 
extracted from DFT calculations by Guo \textit{et al.}~\cite{guo2015first} for each axis. 
However, anisotropic renormalized effective ODP constants ($\beta$) have not been found in the 
literature, to the best of our knowledge. In this case, $\beta$ values 
have been obtained by fitting our transport data 
for $\sigma$ at $\num{300}$~K to the experimental data~\cite{zhao2014ultralow}. 
Hence, the effective ODP's can be considered, 
as emphasized by Conwell~\cite{conwell1983high}, as effective couplings 
representing the actual coupling matrix elements
that account for the more complex ${\bf{q}}$ dependence. The resulting
deformation potentials, both ADP and ODP constants, derived at $\num{300}$~K, 
were directly applied throughout the full range
of temperatures, extending up to $\sim\num{810}$~K.
It is also important to stress that deformation potential constants, 
were considered to be the same in both cases, intrinsic and hole-doped SnSe.
The large anisotropy in the derived ODP constants
is directly related to the strong and weak Sn-Se bonds in the crystal structure. 
It has been shown by heat capacity measurements the existence of two characteristic vibrational scales
corresponding to hard and soft substructures~\cite{popuri2017evidence}.
\begin{table}
	\caption{Parameters used to calculate the RTs of \ce{SnSe} in the \textit{Pnma} phase.
		The parameters are $\zeta_0$ (static dielectric constant)\cite{chandrasekhar1977infrared}, 
                $\zeta_{\infty}$ (high-frequency dielectric constant)\cite{chandrasekhar1977infrared},
		$v_0$ (sound velocity)\cite{guo2015first}, $\rho$ (mass density), $\beta$ 
		(ratio of the average optical phonon energy and optical deformation potential constant), 
                $E_1$ (acoustic deformation potential)\cite{guo2015first}, 
		Gap energy\cite{zhao2014ultralow} and $K_{pol}$ (proportionality factor
		of Fr{\"o}lich interaction to account for the  conduction of p-like symmetry holes).}
	\label{parameters}
	\centering
	\begin{tabular} {llccc}\hline
		Parameter    & a axis & b axis & c axis \\ \hline
		$\zeta_0$  & $47$  & $50$ & $56$  \\
		$\zeta_{\infty}$  & $18$  & $15$ & $15$ \\
		$v_0$~(m/s) & $3356$  & $3267$  &  $3493$ \\
		$\rho$~(Kg/m$^{3}$) & $6180$  & $6180$  &  $6180$ \\
		$E_1$~(eV) & $14.1$  & $15.8$  &  $16.4$ \\
		$\beta$ & $0.000082$  & $0.001702$  &  $0.000136$ \\
		$Gap$~(eV) & $0.86$  & $0.86$  &  $0.86$ \\
		Lattice constant~(\AA) & $11.564$  & $4.542$  &  $4.166$ \\
		$K_{pol}$& $6.2$  & $6.2$  &  $6.2$ \\ \hline
	\end{tabular}
\end{table}

As expected, DFT calculations with PBE functional 
underestimate the band gaps. Consequently,
for the purpose of comparison with experimental data, the calculations of the
TE properties of SnSe were done by rigidly shifting upward the DFT-PBE conduction bands
in order to attain the experimental value of the band gap \num{0.86} eV~\cite{zhao2014ultralow}. 
As will be discussed, the polar mode scattering by optical phonons 
dominate along $b$ axis, consequently, we have adjusted the $K_{pol}$ proportionality factor 
of the Fr{\"o}lich interaction to account for the conduction of p-like symmetry holes, 
which has been considered the same for all axes. 

\subsubsection{Main Scattering Mechanisms and RTs}
\begin{figure*}
\centering
\includegraphics[width=0.95\textwidth]{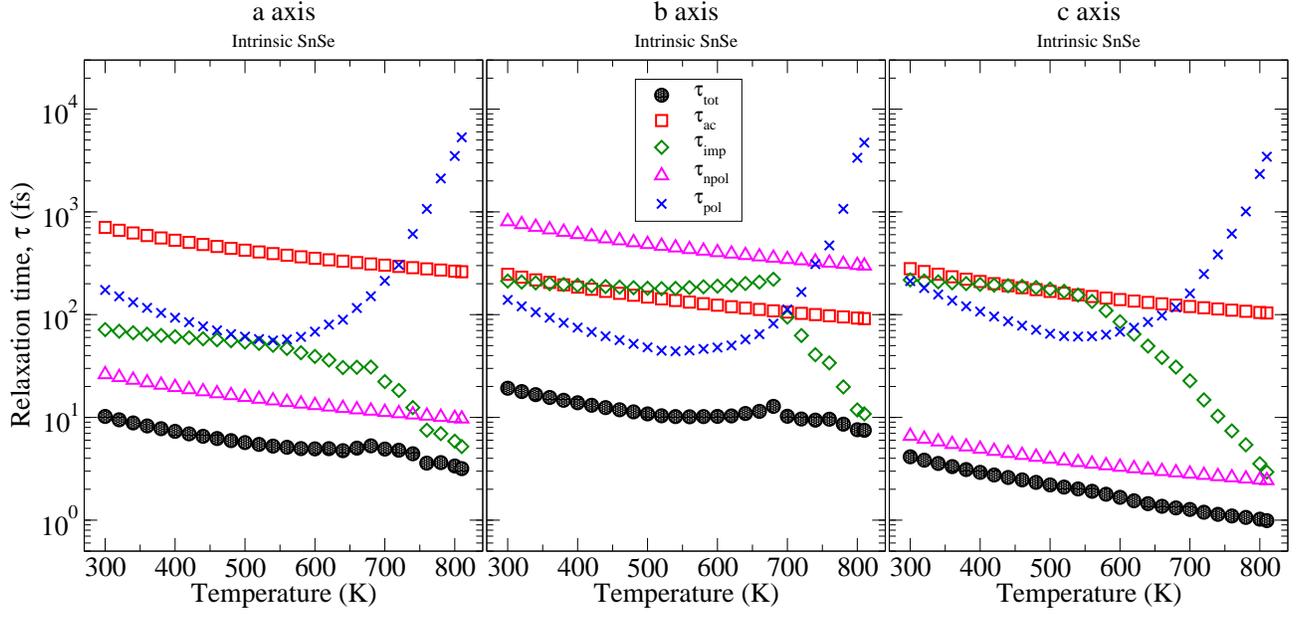}
\caption{Temperature and axial dependence of average RTs 
calculated for each scattering process in intrinsic \ce{SnSe}, 
namely, non-polar scattering of acoustic phonons ($\tau_{ac}$), 
scattering by ionized impurities ($\tau_{imp}$), non-polar scattering of 
optical phonons ($\tau_{npol}$) and polar scattering of optical phonons ($\tau_{pol}$). 
The average total RTs ($\tau_{tot}$) that are calculated from the Mathiessen's rule 
are also shown as a function of temperature for each crystallographic axis.}
\label{relax_time_1}
\end{figure*}

\begin{figure*}
\centering
\includegraphics[width=0.95\textwidth]{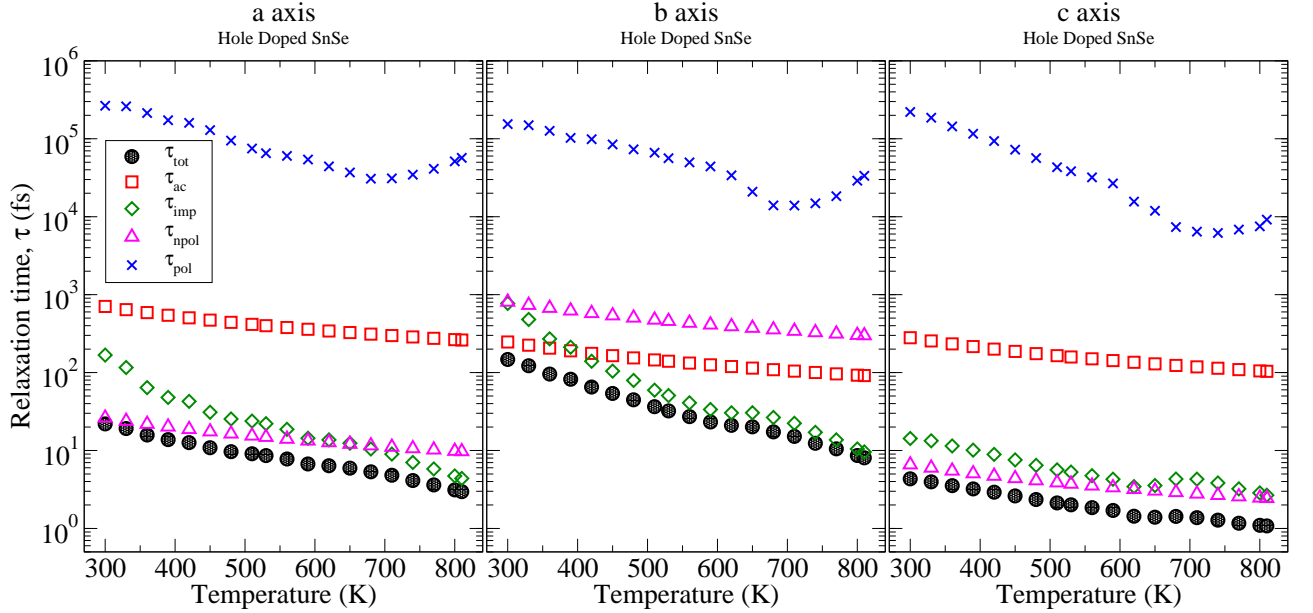}
\caption{Temperature and axial dependence of average RTs
calculated for each scattering process in hole-doped \ce{SnSe},
namely, non-polar scattering of acoustic phonons ($\tau_{ac}$),
scattering by ionized impurities ($\tau_{imp}$), non-polar scattering of
optical phonons ($\tau_{npol}$) and polar scattering of optical phonons ($\tau_{pol}$).
The average total RTs ($\tau_{tot}$) that are calculated from the Mathiessen's rule
are also shown as a function of temperature for each crystallographic axis.}
\label{relax_time_2}
\end{figure*}

Finally, a picture of the axial dependence of each scattering processes with temperature 
and chemical potential can be inferred from the average of each RT 
over the band indices $n$ and the wavevectors $\bf{k}$,  
\begin{equation}
(\tau_{x})_{t}(\mu_{t},T) = \frac{1}{N_k{\cdot}N_{band}}\sum_{n}\sum_{k}(\tau_{x})_{t}(n,k,\mu_{t},T)~,
\end{equation}
where $N_k$ is the number of $\bf{k}$-points sampled, $N_{band}$ is the number of bands, and 
$x$ stands for the charge carrier scattering mechanism, namely, non-polar scattering by acoustic phonons ($\tau_{ac}$), 
nonpolar scattering by optical phonons ($\tau_{npol}$), 
polar scattering by optical phonons ($\tau_{pol}$) and scattering by ionized impurities ($\tau_{imp}$).  
The total average RT ($\tau_{tot}$) is calculated through the Mathiessen's rule.
The calculated average RTs
as functions of the temperature for each axis of intrinsic and hole-doped \ce{SnSe} are shown in 
Figs.~\ref{relax_time_1} and~\ref{relax_time_2}, respectively.
As it will be discussed below, in general, the dominant scattering mechanism varies with temperature and 
is not the same for each axis, highlighting the 
strong SnSe anisotropy. 

The scattering by acoustic phonons strengthens with temperature,
for both intrinsic and hole-doped SnSe, this is
intimately related to the enhancement in the number of available phonons 
for scattering as temperature increases. Figs.~\ref{relax_time_1} and \ref{relax_time_2} show that the RT, $\tau_{ac}$, associated with this process, decreases as $1/T$.
In the $b$ axis, the scattering by acoustic phonons 
plays a more proeminent role than in the other axes, 
highlighting the strong anisotropy, however, this kind of scattering mechanism 
is only secondary in the overall transport process.
 
For the intrinsic case, the polar scattering by optical phonons strengthens as temperature increases, reducing $\tau_{pol}$ approximately 
as $1/T$ up to $\sim\num{550}$~K. 
Above that temperature, this scattering process begins to weaken, consequently increasing $\tau_{pol}$. This behavior is 
closely related to the onset of the thermally activated process
of vacancies formation, which effectively increases the number of charge carriers in the material, 
hence inducing a stronger screening of the polarization. 
The onset of this screening effect on the polar scattering by optical phonons occurs 
in all axes at around the same temperature, however, this behavior mostly affects 
the overall transport in the $a$ and $b$ axes, in which $\tau_{tot}$ tends to 
increase following $\tau_{pol}$, despite the competition 
with the other scattering processes.  

The RT due to nonpolar scattering by optical phonons, $\tau_{npol}$, also decreases as $1/T$, 
for both intrinsic and hole-doped SnSe. 
As discussed in the previous section, 
the strong anisotropy of SnSe is clearly shown by the fitted ODP constants $ E_0=\hbar\omega_0/\beta$. Even though, the ODP for 
the $a$ axis is larger than that for the $c$ axis, the resulting RTs is 
relatively smaller than that for the $c$ axis, emphasizing the importance of 
the band velocities in the calculation of RTs.
 

The RT due to scattering by ionized  impurities in the intrinsic case is almost constant up to 
temperatures at which the process of thermal creation of vacancies becomes relevant with respect
to preexisting vacancies created during growth. Above those temperatures, the
concentrations of vacancies begin to increase rather fast, and the RT associated with this scattering 
mechanism drops quite rapidly as well, for all three axes.


For hole-doped \ce{SnSe}, the dominant scattering mechanisms
are the nonpolar scattering by optical phonons and scattering by charged vacancies.
The latter mechanism dominates along the $b$ axis for the
whole temperature range and becomes the primary scattering mechanism
above $~700$~K for the $a$ axis, while strongly
affects the overall transport properties for the $c$ axis
at higher temperatures. It reflects the strong
influence of the thermally activated
process of vacancy formation for TE transport properties at higher
temperatures, especially above $~700$~K.
In fact, in the case of hole-doped \ce{SnSe}, in which the holes
concentration is large and the screening is 
stronger, the polar scattering by optical phonons
pratically does not cause pronounced effects on the transport
properties, since $\tau_{pol}\sim10^{-11}\:\text{s}$. 

\subsubsection{Electrical Conductivity}
\begin{figure}
\centering
\includegraphics[width=0.45\textwidth]{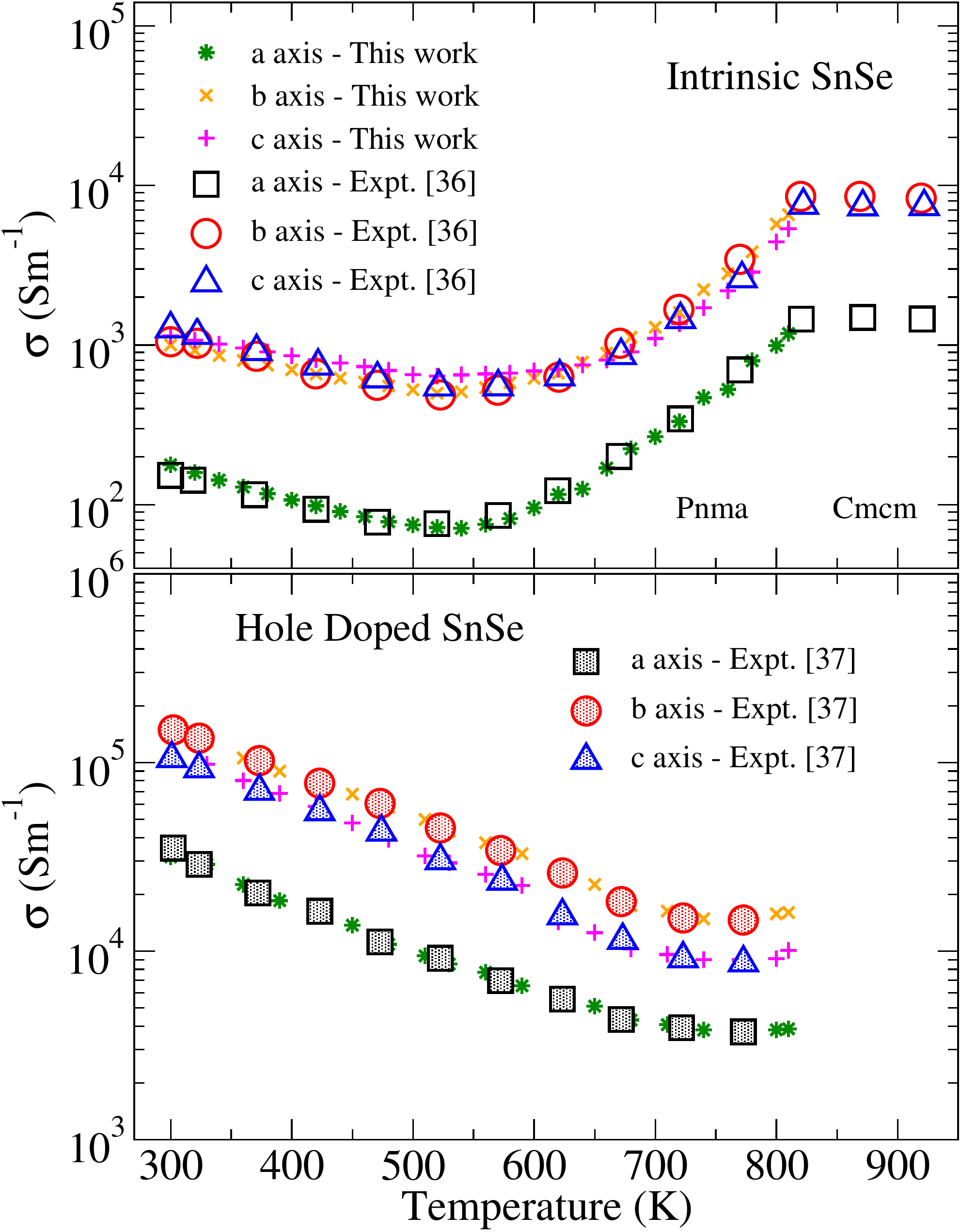}
\caption{\ce{SnSe} anisotropic electronic conductivity, $\sigma$, as a function of temperature, 
for intrinsic \ce{SnSe} (top panel) and for hole-doped \ce{SnSe} (bottom panel). 
Our findings are plotted along with experimental results~\cite{zhao2014ultralow,zhao2015ultrahigh}.}
\label{elec_cond}
\end{figure}
Our results for the anisotropic electronic conductivity, $\sigma$, are in close agreement 
with experimental data\cite{zhao2014ultralow,zhao2015ultrahigh} (see Fig.~\ref{elec_cond}).
It is noteworthy to point out that, at first glance, this good agreement would result solely from
the fitting procedure used to obtain the concentration of ionized impurities, $N_i$, in the relaxation time $\tau_{imp}$.
This is not the case, because the electrical conductivity also depends on the chemical potential and
carrier concentration, which are obtained from the fitting of the calculated Seebeck coefficient to its experimental data. 
This indicates a consistency between the calculated Seebeck coefficient and electrical conductivity.  
The magnitude of $\sigma$ along the $a$ axis is almost one decade smaller than that for $b$ and $c$ axes. 
This is related to the strong anisotropy of the chemical bonding. 

For intrinsic \ce{SnSe}, between $300-550$~K, $\sigma$ decreases with temperature as a direct consequence of the 
enhancement of the scattering processes with temperature, while the hole concentration is maintained approximately constant 
in that temperature range. Above $\sim550$~K, the thermally activated process of defects formation sets in 
and, consequently, $\sigma$ markedly increases with temperature. This behavior is attained even though the 
scattering by charged vacancies becomes more relevant, however, the enhancement in the screening 
of the polar scattering by optical phonons largely  
compensates for the effect of scattering by ionized impurities.   
For the $b$ axis the polar scattering by optical phonons dominates. 
Consequently, the overall scattering 
is strongly influenced by this mechanism. 
As the temperature increases and the activated process 
of vacancy formation sets in, the generation of more 
free carriers causes a stronger screening in this 
scattering mechanism, consequently increasing the 
electrical conductivity in the $b$ axis to 
values even larger than those for the $c$ axis for temperatures above $\sim\num{550}$~K, 
where the mechanism of polar optical phonons scattering is only secondary. 

In the case of hole-doped SnSe, $\sigma$ presents a metal-like 
behavior, in which $\sigma$ decreases steadily  with temperature, up to 
high temperatures ($\sim800$~K), differently than what is observed in the intrinsic case. Above this temperature, 
$\sigma$ tends to increase as a consequence of the pronounced 
enhancement in the hole concentration with temperature. 
In fact, this pronounced enhancement in the hole 
concentration begins around $\num{670}$~K. However, unlike the intrinsic case,
where $\sigma$ begins to increase, for hole-doped \ce{SnSe}, $\sigma$ continues decreasing,  
highlighting the stronger influence of the scattering by ionized charged impurities. 

\subsubsection{Seebeck Coefficient}
\begin{figure}
\centering
\includegraphics[width=0.45\textwidth]{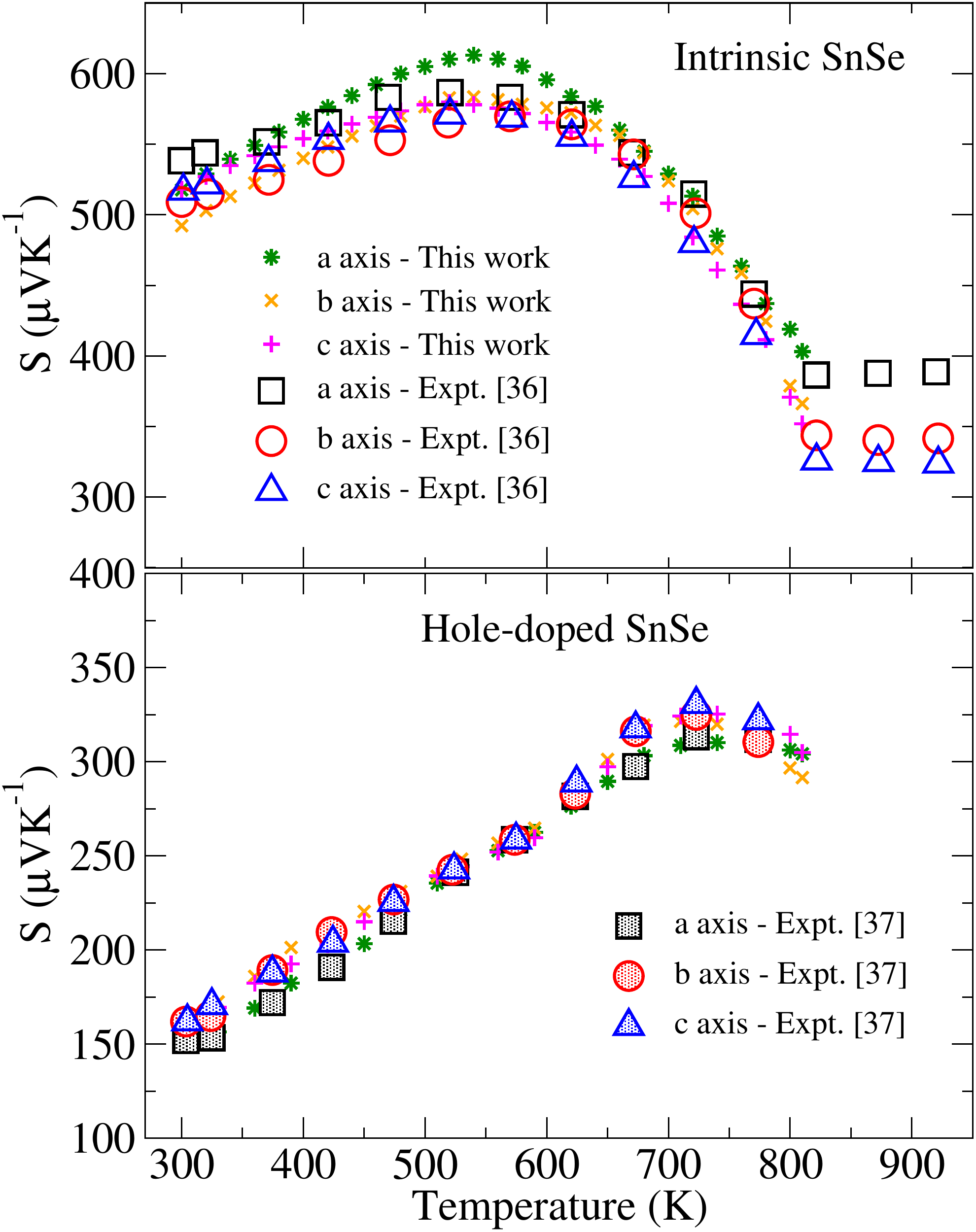}
\caption{\ce{SnSe} anisotropic Seebeck coefficient, $S$, as a function of temperature 
for intrinsic \ce{SnSe} (top panel) and for hole-doped \ce{SnSe} (bottom panel). 
Our findings are plotted along with experimental results~\cite{zhao2014ultralow,zhao2015ultrahigh}.}
\label{seebeck}
\end{figure}
Our results for the Seebeck coefficients, $S$, as a function of temperature, are shown in Fig.~\ref{seebeck} 
together with experimental results for both intrinsic \ce{SnSe}~\cite{zhao2014ultralow} and hole 
doped \ce{SnSe}~\cite{zhao2015ultrahigh}. The close agreement between calculations and experiment results from two facts: (1) 
the evolution of hole concentration with temperature was taken into consideration and (2) the relaxation times obtained from the
electrical conductivity calculations are properly described. This demonstrates again the consistency between the calculated
Seebeck coefficient and electrical conductivity.
At $\sim550$~K, the experimental Seebeck coefficients of intrinsic \ce{SnSe} exhibit unpronounced maxima, 
which can be attributed mainly to the increase of hole concentration. A similar maximum 
occurs to the hole-doped \ce{SnSe}, however at higher temperatures ($\sim750$~K).

\subsubsection{Carrier Thermal Conductivity}

 \begin{figure}
        \centering
        \includegraphics[width=0.45\textwidth]{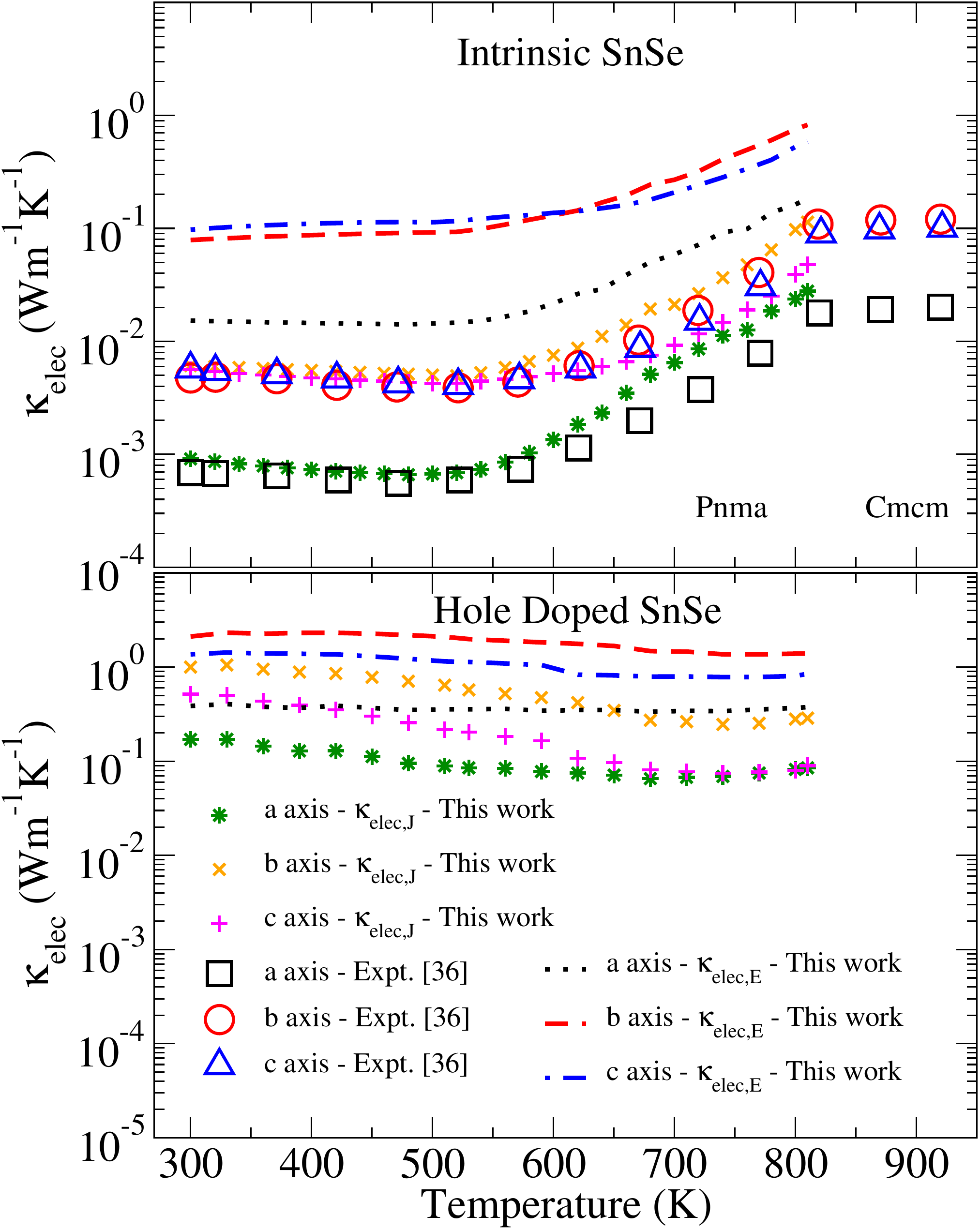}
        \caption{\ce{SnSe} anisotropic thermal conductivities due to the carriers transport, $\kappa_{elec,J}$ and $\kappa_{elec,E}$, as functions of temperature for intrinsic \ce{SnSe} (top panel) and for hole-doped \ce{SnSe} (bottom panel). 
                 Our findings are plotted along with available experimental results for $\kappa_{elec,J}$~\cite{zhao2014ultralow}.}
        \label{Kl}
 \end{figure}

Our results for the thermal conductivity due to the carrier transport are shown in Fig.~\ref{Kl}, for both cases, intrinsic (top panel)
and hole-doped (bottom panel) SnSe. 
These results are related to two dissimilar experimental situations, namely, $\kappa_{elec,J}$ and $\kappa_{elec,E}$. 
The former corresponds to the electric open-circuit condition where the electric current throughout the system 
is null. Experimentally, this situation corresponds to the measurement process performed 
by a device with high internal electrical resistance. 
The latter situation corresponds to the electric short-circuit condition where the electrochemical 
potential is constant throughout the system. Both experimental conditions are related by 
$\kappa_{elec,J} = \kappa_{elec,E} - T\sigma S^2$, resulting that $\kappa_{elec,E} > \kappa_{elec,J}$, 
as observed from Fig.~\ref{Kl}. 
Moreover, the thermal conductivity due to the carriers 
is greater in the hole-doped system compared to the intrinsic case. 

The Lorenz number, $L$, defined as 
\begin{equation}
L = \frac{\kappa_{elec,J}}{\sigma}T~,
\end{equation}
is commonly considered as a constant in order to determine $\kappa_{elec,J}$, a practice 
that is extensively used. Zhao {\it{et. al}}~\cite{zhao2014ultralow} have considered 
the non-degenerate limit, where $L = 1.5\times10^{-8}\rm{V}^2\rm{K}^{-2}$, to obtain $\kappa_{elec,J}$ for SnSe.
Our findings for $\kappa_{elec,J}$ can be compared with 
Zhao's results for intrinsic SnSe in Fig.~\ref{Kl} (top panel). One can observe
an excellent agreement for the three axis for temperatures $\lesssim600$~K.
It should be emphasized that our results for $\kappa_{elec,J}$ are completely free from any empirical fitting procedure.
However, above $\sim600$~K, for the $a$ and $b$ axes our results are greater 
than Zhao's results, which were determined considering $L$ as a constant, 
while for the $c$ axis our values are smaller than the experiment.
Our results suggest the fact that $L$ cannot be considered as a constant. 
Since SnSe has a complex band structure and the RTs have a complex
behavior with the electron energy, $L$ dependence on
temperature and chemical potential should be taken into account.   
Such deviations between our results and Zhao's results may have significant implications,
resulting in corrections to the earlier predicted values of the 
figure of merit~\cite{zhao2014ultralow}, $zT$, which may have been 
overestimated for $a$ and $b$ axes and underestimated for $c$ axis in the intrinsic case. 
Fig.~\ref{Kl} also depicts our results for $\kappa_{elec,E}$ for both cases intrinsic and hole-doped SnSe. 
It is important to note that in both cases $\kappa_{elec,E}$ is larger than 
$\kappa_{elec,J}$. The difference between these two thermal conductivities can be up one order of magnitude
for the intrinsic case. This is a manifestation of the large power factor of SnSe.

\begin{figure}
        \centering
        \includegraphics[width=0.45\textwidth]{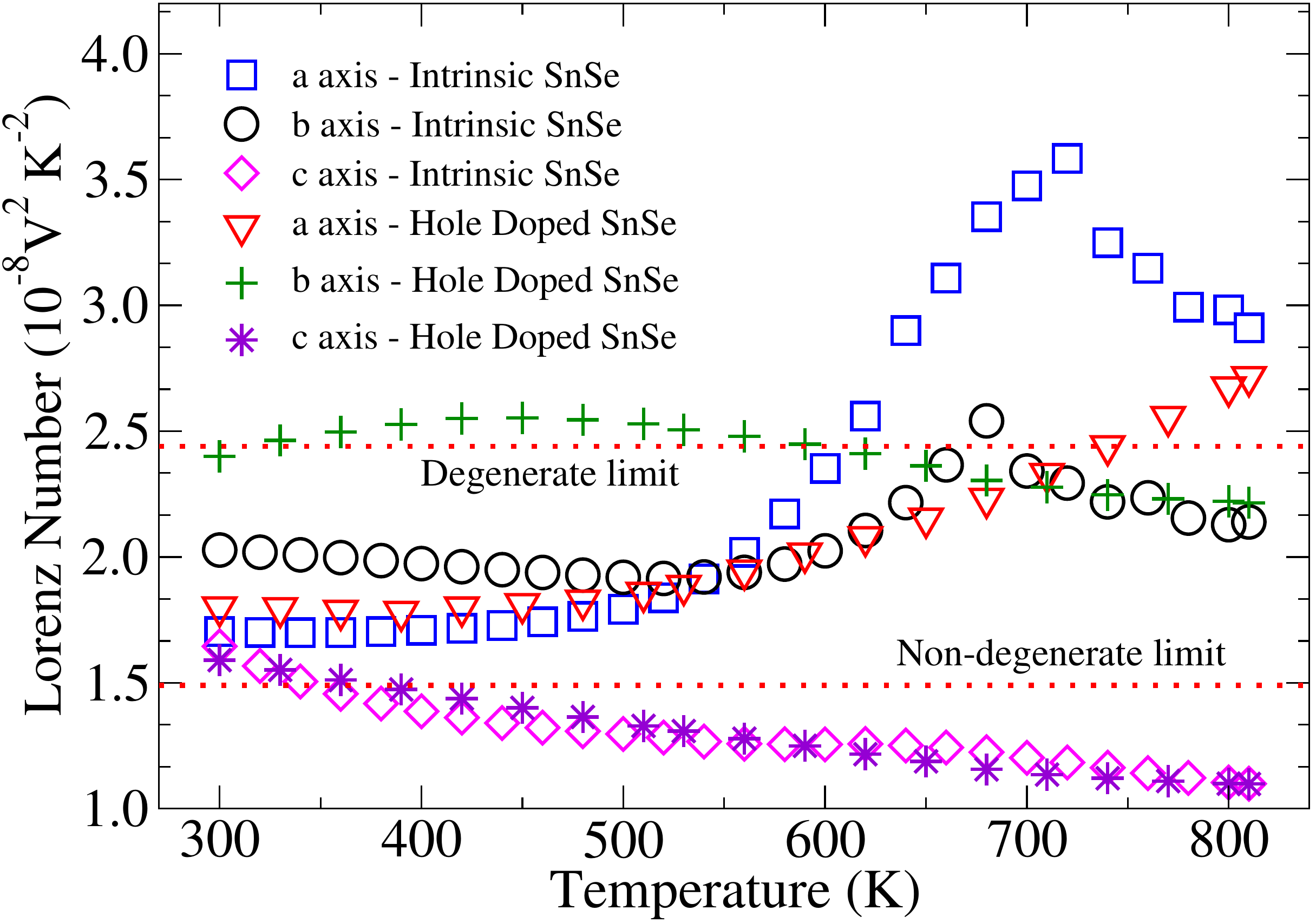}
        \caption{\ce{SnSe} anisotropic Lorenz number, $L$, as a function of temperature
                for intrinsic \ce{SnSe} and for hole-doped \ce{SnSe}. The horizontal lines indicate the non-degenerate limit of 
                semiconductors ($1.5\times10^{-8}V^2K^{-2}$) and the degenerate limit ($2.44\times10^{-8}V^2K^{-2}$) defined 
                by the Wiedemann-Franz law.} 
        \label{Lorenz}
 \end{figure}

Fig.~\ref{Lorenz} shows the results of our calculations for anisotropic $L$. 
At $300$~K, $L$ values are within the non-degenerate and degenerate limits, being the latter 
defined by the Wiedemann-Franz law~\cite{wang2018calculation}. As temperature increases, we can 
properly observe the variation of $L$ for each crystallographic axis 
of both hole-doped and intrinsic SnSe. There is a large discrepancy between the results depending on the 
crystallographic axis. In general, $L_c$ values for $c$ 
axis tend to be around the non-degenerate limit or below, while for $a$ and $b$ axes the values are much larger. 
The ordering is basically $L_c~<~L_a~<~L_b$ at low temperatures, 
with an inversion between $a$ and $b$ axes as temperature increases. For the 
intrinsic case, above $\sim550$~K, which is the  onset temperature for vacancies formation, 
$L_a$ begins to be greater than $L_b$. Similarly, such inversion occurs at a higher temperature, 
$\sim700$~K, for the hole-doped case and can be attributed to a fast increase of $\kappa_{elec,J}$ 
for the $a$ axis with temperature, which is not evident from Fig.~\ref{Kl} because of the logarithmic scale. 
At $810$~K, $L$ assumes very similar values for each one of the axes, for both cases hole-doped and intrinsic SnSe, 
because the carrier concentrations tend to be closer in both cases. 

\subsection{Summary}

Based on the solution of the linearized BTE and using 
smooth Fourier interpolation of the KS eigenvalues and their corresponding velocities,  
we propose a methodology that incorporates models of anisotropic RTs for the 
main scattering mechanisms to calculate 
TE transport coefficients within the \texttt{BoltzTraP} code. 
The scattering mechanisms considered include the non-polar scattering 
by acoustic phonons (as described within the deformation potential theory), 
screened polar (based on the Fr{\"o}lich theory) and non-polar scattering (also 
within the deformation potential theory)
by optical phonons and the scattering by ionized impurities with screening
(within the Brooks-Herring approach).  
Such models for RTs  
include dependence on temperature, 
chemical potential and \textit{ab initio} bandstructure, consequently, the methodology is able to directly capture 
important contributions to TE properties from multiple 
band edges, degenerate band edges and non-parabolicity effects on the same footing. 

Therefore, our implementation allows to directly grasp and 
understand the physical mechanisms underlying 
the experimental data, including the knowledge of main scattering processes and 
their evolution with temperature and chemical potential. This methodology was applied to both
intrinsic and hole-doped SnSe. 
For intrinsic \ce{SnSe}, within the range $300-550$~K, 
in which the hole concentration is approximately constant, 
$\sigma$ decreases with temperature as a direct consequence of the
enhancement of the scattering processes with temperature. 
Above $\sim550$~K, the thermally activated process of defects formation sets in
and, consequently, $\sigma$ markedly increases with temperature, even though the
scattering by charged vacancies becomes more relevant. However, the enhancement in the screening
of the polar scattering by optical phonons largely
compensates for the effect of scattering by ionized impurities.
In the case of hole-doped SnSe, $\sigma$ presents a metal-like
behavior, with $\sigma$ decreasing steadily as temperature increases, up to
high temperatures ($\sim800$~K), highlighting the stronger influence 
of the scattering by ionized charged impurities.

Furthermore, toward transport properties optimization for TE applications,
the methodology presented here can be extended to strained materials considering
almost the same constants of unstrained systems. Particularly,
this is a motivation for strained SnSe due to the sensitivity of its
band-edge states to lattice strains~\cite{wu2017engineering}.
Most importantly, the calculations take approximately the same amount of
computational cost as the calculation using constant RT, as developed
in the public version of \texttt{BoltzTraP} code.
We expect that our approach can be applied to a broad range 
of semiconducting materials and potentially investigate  
the enhancement of TE properties in next-generation TE materials that 
explore materials anisotropy with multiplicity, 
degeneracy and non-parabolicity of the band edges. 
 
%

\section*{Acknowledgements}

ASC and AA gratefully acknowledge support from the Brazilian agencies CNPq and FAPESP under 
Grants \#2010/16970-0, \#2013/08293-7, \#2015/26434-2, \#2016/23891-6, \#2017/26105-4, and \#2019/26088-8. 
JJM acknowledges funding by the Ministerio de Econom{\'i}a y Competitividad, Agencia Estatal de Investigaci{\'o}n, 
and Fondo Europeo de Desarrollo Regional (FEDER) (Spain and European Union) through Grant No. PGC2018-094763-B-I00, 
and by Junta de Extremadura (Spain) through Grant No. IB16013 (partially funded by FEDER). 
The calculations were performed at CCJDR-IFGW-UNICAMP in Brazil.


\section*{Computer Code Availability}
All computer implementations of the methodology developed in this project were written 
in Fortran 90 and are available upon reasonable request.

\bibliographystyle{apsrev4-1}
{\footnotesize
\bibliography{Ref2.bib}}

\end{document}